\documentclass[11pt]{article}
\usepackage[OT1]{fontenc}
\usepackage{epsfig}
\usepackage{amsmath}
\usepackage{amssymb}
\usepackage{natbib}
\usepackage{siunitx}

\title{X-ray Halos around Massive Galaxies: \\ Data and Theory}
\author{\'Akos Bogd\'an and Mark Vogelsberger}
\date{}
\begin{document}

\maketitle

\section*{Abstract}
The presence of gaseous X-ray halos around massive galaxies is a basic prediction of all past and modern structure formation simulations. The importance of these X-ray halos is further emphasized by the fact that they retain signatures of the physical processes that shape the evolution of galaxies from the highest redshift to the present day. In this review, we overview our current observational and theoretical understanding of hot gaseous X-ray halos around nearby massive galaxies and we also describe the prospects of observing X-ray halos with future instruments.  

\section{Introducing X-ray halos around massive galaxies}

\subsection{Motivation}

The first gaseous X-ray emitting halos around massive galaxies in the Virgo cluster were observed nearly 50 years ago using the \textit{Einstein} Observatory \citep{mathews78,forman79}, but theoretical considerations suggesting the existence of such halos date back even earlier. The existence of an extended gaseous halo around the Milky Way was first hypothesized by \citet{spitzer56}, who suggested that this circumgalactic medium (henceforward CGM) would have a temperature of $10^6$~K and an electron density of $5\times10^{-4} \ \rm{cm^{-3}}$. Based on these properties, the radiation from the CGM is expected to fall in the X-ray waveband. Interestingly, studies performed with modern-day X-ray telescopes, such as the \textit{Chandra} X-ray Observatory confirmed the presence of the CGM around the Milky Way \citep[e.g.][]{gupta17} and established that the X-ray properties of the hot gas are, in fact, comparable to those suggested by \citet{spitzer56}.   
Although the initial prediction by \citet{spitzer56} only forecasted the presence of an X-ray emitting CGM around the Milky Way, about two decades later, \citet{white78} generalized this picture and suggested that luminous X-ray halos are an essential part of galaxy formation. They proposed a two-stage theoretical picture of galaxy formation. In the first stage, the baryonic material originating from the intergalactic medium is accreted onto the dark matter halos of galaxies. During the infall, accretion shocks heat the gas to the virial temperature of the galaxies, which -- for massive dark matter halos -- is a few million Kelvin. Therefore, the X-ray halos of massive galaxies are expected to emit in the X-ray waveband. In the second stage of galaxy formation, the radiative cooling of the hot X-ray emitting gas results in cooling flows, which provide material for star formation, thereby building up the stellar content of galaxies.  

This initial model of galaxy formation was further improved by \citet{white91}, who suggested that the X-ray emitting CGM can extend beyond the stellar body of galaxies and they are ubiquitous around \textit{all} Milky Way-type (i.e.\ disk-dominated) galaxies. They also established that the X-ray gas cools via line emission and thermal bremsstrahlung and provides material to the formation of stars. These processes were shown to be essential to reproduce the observed diversity of galaxy morphology, the disruption of disks due to mergers, and bar instabilities. Overall, \citet{white91} demonstrated that the CGM is an integral part of galaxies, and the physical processes associated with its heating and cooling play a profound effect on the overall evolution of galaxies. Therefore, to understand the evolution of galaxies from high redshifts to the present epochs, it is indispensable to model the evolution of the CGM. In fact, comparing the theoretical predictions of the CGM with its observable properties allows to probe and constrain a multitude of complex physical processes, which are not easily accessible by other observational studies. 

Because observations of the CGM around present-day galaxies can provide a wealth of information about the origin and evolution of galaxies, it is important to investigate the basic characteristics of the CGM. The infalling gas is expected to be shock-heated to the virial temperature of the host galaxy's dark matter halo, which temperature can be described as: 
\begin{equation}
    T_{\rm 200} = \frac{1}{3}\frac{\mu m_{\rm p}}{k_{\rm B}} v_{\rm 200}^2= 3.6\times10^5 \ \rm{K} \Big( \frac{\mu}{0.59} \Big) \Big( \frac{v_{\rm 200}}{100 \ \rm{km \ s^{-1}}} \Big)^2 \ ,
\end{equation}
where $\mu$ is the mean molecular weight, $m_{\rm p}$ is the proton mass, $k_{\rm B}$ is the Boltzmann constant, and $v_{\rm 200}$ is the circular velocity of the galaxy. For massive galaxies that exhibit $v_{\rm 200} \gtrsim 200 \ \rm{km \ s^{-1}}$, this characteristic temperature will be a few million Kelvin (or $\gtrsim0.1$~keV) and the gas will emit in the soft X-ray regime. Because the cooling time of the X-ray gas around massive galaxies exceeds the dynamical time, the X-ray halos will be quasi-static on timescales comparable to the Hubble-time. This directly implies that the infalling gas that was heated to X-ray temperatures at the formation of the galaxies should remain observable even in the present-day Universe as the CGM. During the evolution of galaxies, the characteristics of the CGM are expected to change due to various physical processes. Thus, the large-scale CGM reflects both the basic principles of galaxy formation and retains a memory of the physical processes that shaped the galaxies throughout their evolution. Therefore, the CGM of galaxies provides a superb observational probe to simultaneously constrain both the basic principles of galaxy formation and the high-level physics of galaxy evolution.

\subsection{Overview of past X-ray observations}
\label{sec:overview}

Recognizing the importance of the CGM around massive galaxies, a large number of observational campaigns were carried out in the past decades to detect the hot gaseous halos and characterize their properties. In the following subsections, we overview the main observational efforts to explore the CGM around massive elliptical and disk-dominated galaxies.  

\subsubsection{Massive elliptical galaxies}
\label{sec:overview_elliptical}

In elliptical galaxies, the hot X-ray emitting gas can have two different origins. First, similar to disk-dominated galaxies, part of it may arise from the infall of the primordial gas onto the dark matter halos, which gas was shock-heated to the virial temperature of the dark matter halo. Second, the gas ejected from evolved stars and planetary nebulae may collide with ambient gas and gets shock heated to the kinetic temperature of the galaxy, which is equivalent with the galaxy's stellar velocity dispersion. However, the relative contribution of these different sources to the overall emission has remained unclear. 

The existence of luminous X-ray halos around elliptical galaxies was established based on \textit{Einstein} and \textit{ROSAT} X-ray observations \citep{forman85,trinchieri85,mathews90,mathews03}. These studies demonstrated the ubiquitous nature of extended X-ray halos around elliptical galaxies and showed that these halos often extend beyond the galaxy's stellar distribution. While the luminous nature of the X-ray emission has been established around ellipticals, the  relative contribution of different types of sources and the physical state of the hot gas has been debated. This was drastically changed with the launch of \textit{Chandra} and \textit{XMM-Newton}, which instruments provided a definite edge on previous instruments. Specifically, the high angular resolution of these instruments allowed to resolve bright point sources in the galaxies, such as low-mass X-ray binaries or AGN \citep{gilfanov04,fabbiano06}. In addition, detailed studies of the Milky Way and some nearby galaxies, e.g.\ M32 or NGC3379, allowed detailed studies of even fainter X-ray stellar sources, such as active binaries and cataclysmic variables and to estimate the contribution of these unresolved point sources to the integrated emission of the galaxies \citep{sazonov06,revnivtsev06,revnivtsev08}. 

Since X-ray halos were routinely detected around massive elliptical galaxies, many studies explored the most fundamental properties of the gas, namely its X-ray luminosity and gas temperature. Since a substantial number of galaxies were observed, a statistically significant sample could be studied and basic correlations could be established. Specifically, the luminosity of the gas correlates with the gas temperature for all virialized systems, including galaxy clusters, galaxy groups, and galaxies. However, the slope of the ${L}_{{\rm{X,gas}}} - {T}_{{\rm{gas}}}$ relation exhibits differences: While galaxy clusters exhibit  ${L}_{{\rm{X,gas}}}\propto {T}_{{\rm{gas}}}^{2-3}$, galaxy groups have a steeper relation with ${L}_{{\rm{X,gas}}}\propto {T}_{{\rm{gas}}}^{3-4}$, on galaxy scales the relation may become even steeper, ${L}_{{\rm{X,gas}}}\propto {T}_{{\rm{gas}}}^{4.5}$. The breakdown in the scaling relation in galaxies is likely due to the baryon physics, such as the feedback from AGN and supernovae, which play a more significant role than gravity. Additionally, it was established that the X-ray luminosity of elliptical galaxies exhibits a correlation with the optical properties of galaxies. Specifically, \citet{osullivan01} concluded that the X-ray luminosity of the hot gas approximately correlates with the B-band luminosity of galaxies. Since the B-band mass-to-light ratios exhibit large scatter, follow-up studies used the K-band luminosity and/or the stellar mass of galaxies to establish trends between the X-ray luminosity of gaseous halos and the physical properties of galaxies. These studies established that there is a correlation between these properties, but the scatter in the ${L}_{{\rm{X,gas}}}\mbox{--}{M}_{\star}$ relation is larger than expected based on a simple expectation. The large scatter can be attributed to physical processes that influence the physical properties of the gas, notably the energetic feedback of the AGN and the rotation properties of the galaxies. To further explore these scaling relations, \citet{kim13} probed the relation between the X-ray luminosity and the total mass of galaxies within 5 effective radii ($M_{\rm{tot}}$). They found a tight correlation for galaxies in the ${L}_{{\rm{X,gas}}} = 10^{38}-10^{43} \rm{erg \ s^{-1} }$ luminosity range with an rms deviation of a factor of three. Surprisingly, this correlation is much tighter than that obtained between the X-ray luminosity and the stellar mass. This hinted that the most important factor in regulating the luminosity and amount of gas may be the total gravitating mass of the galaxy. As a caveat, it needs to be considered that this work is based on a relatively small sample of galaxies and the total mass measurements only extend out to 5 effective radii, which corresponds to $3-30$~kpc, hence are a small fraction of the virial radius of galaxies, implying that only a few percent of the total gravitating mass is included in these regions. 

To establish the relations for large samples, volume-limited surveys were used. The ${\mathrm{ATLAS}}^{{\rm{3D}}}$ sample explored all nearby early-type galaxies with ${M}_{K}< -21.5$ mag. Due to the nearby nature of the galaxies in this survey, this survey contains very few massive ${M}_{\star}> {10}^{11}{M}_{\odot }$ galaxies. They found evidence that slow rotators exhibit lower X-ray luminosity at a given stellar mass than fast rotators. To explore more massive galaxies, the volume-limited MASSIVE survey focused on the most massive nearby galaxies in the local ($D<100$~Mpc) universe. They established a universal scaling law such that ${L}_{{\rm{X,gas}}}\propto {T}_{{\rm{gas}}}^{4.5}$ and demonstrated that the scatter in the X-ray luminosity around the stellar mass is driven by the temperature of the gas. See also the Chapter on the hot ISM in this Section (Nardini, Kim \& Pellegrini). 

To study a large statistical sample of galaxies, which is not feasible using pointing X-ray telescopes, archival \textit{ROSAT} data was used. Because most of these galaxies are not detected individually, the stacking technique has been applied to enhance the signal-to-noise ratios. In the stacking analysis, the X-ray photons associated with many different galaxies are co-added (i.e.\ stacked). While individual galaxies are too faint to be detected above the noise, by co-adding them a statistically significant could be detected since the random noise cancels out, while the signal from the galaxies becomes enhanced. By stacking the ROSAT X-ray data of more than 2000 isolated galaxies, the X-ray emission was detected and they could establish an average luminosity of $\sim10^{40} \ \rm{erg \ s^{-1}}$. Interestingly, they did not find a statistically significant difference between the luminosity of elliptical and spiral galaxies \citep{anderson13}. While the nominal X-ray luminosity of ellipticals is slightly higher, this may be due to the fact that elliptical galaxies are -- on average -- more massive than their spiral counterparts. In a similar study, the X-ray halos of more than 3000 elliptical galaxies were stacked and galaxies with $\sim4\times10^{10} \ \rm{M_{\odot}}$ were detected \citep{bogdan15a}. Since \textit{ROSAT} does not have the angular resolution or the energy resolution to distinguish the emission from hot X-ray halos and the population of unresolved X-ray sources (X-ray binaries), scaling relations were used to estimate the contribution of hot gas, which found that substantial fraction of the emission originates from truly diffuse hot gas.

%According to the theoretical models, today's giant elliptical galaxies formed in massive dark matter halos. It is believed that these galaxies underwent a two-stage galaxy formation process \citep{thomas05,naab09,vandokkum10}. In the first stage, due to the gas infall to the center of dark matter halos, the compact galaxy core built up rapidly. For sufficiently massive halos ($M_{\rm halo} \gtrsim 10^{12} \ \rm{M_{\odot}}$), the cooling time of the infalling gas is longer than the free-fall time, implying that further gas accretion virtually ceases. Because the infalling gas is heated to the virial temperature by accretion shocks, the formation of luminous X-ray halos around elliptical galaxies dates back to the earliest epochs of galaxy formation. During the second stage of the evolution, the galaxies experience numerous minor and major mergers, which contributed to the further growth in mass and size of the galaxies. In addition, these mergers played a major role in the morphological transformation (e.g.\ destroying disks) and contribute to the diversity of the galaxies observed in today's universe \citep{bell04,faber07}. Clearly, the mergers and the associated physical processes (e.g.\ starbursts) can drastically influence the physical properties of the CGM around massive elliptical galaxies. 

\begin{figure}[!tbp]
	\centering
	\vspace{-1cm}
	\begin{minipage}[b]{0.95\textwidth}
		\includegraphics[width=0.48\textwidth]{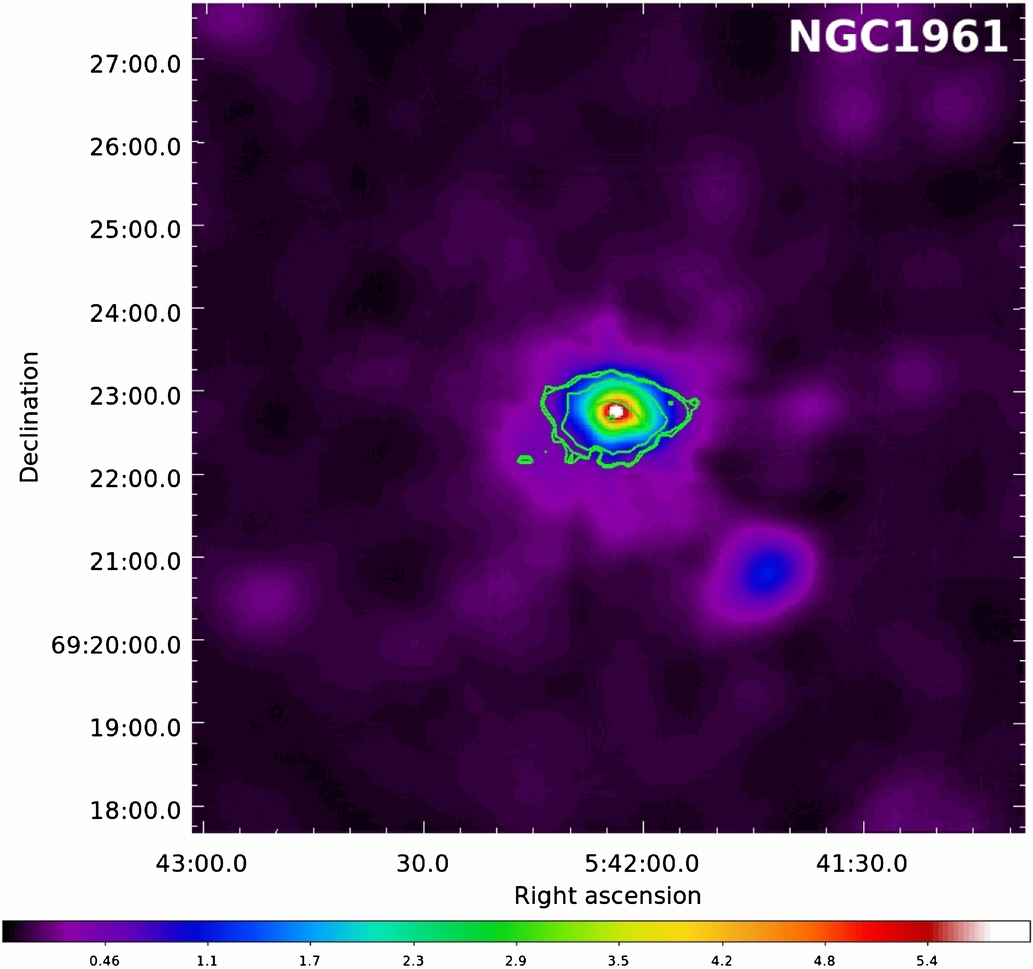}
		\includegraphics[width=0.48\textwidth]{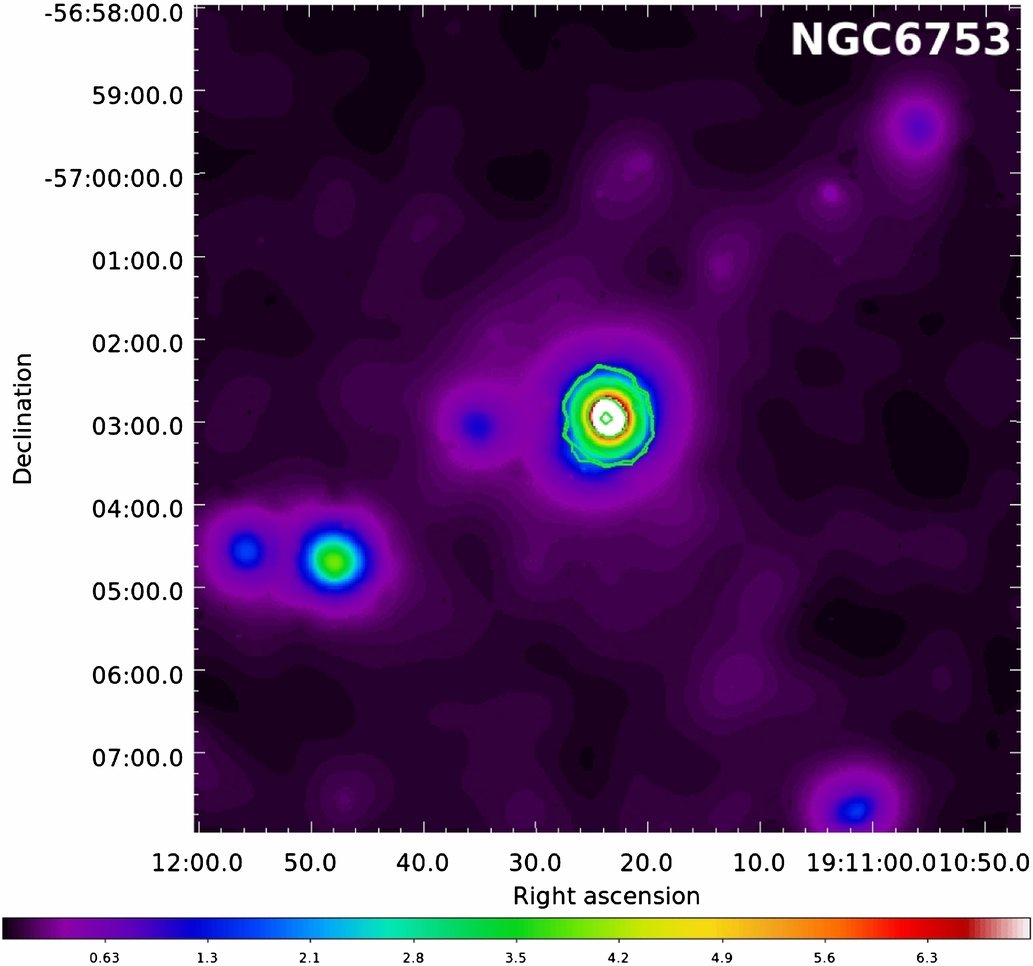}
	\end{minipage}
	\vspace{0cm}
	\caption{Denoised surface brightness image of NGC~1961 (left panel) and NGC~6753 (right) based on \textit{XMM-Newton} observations. The size of the images are $10'\times10'$, which corresponds to $162\times162$~kpc for NGC~1961 and $127\times127$~kpc for NGC6753. These were the first normal (i.e.\ non-starburst) galaxies that were shown to have a luminous CGM. The diffuse X-ray emission extends beyond the stellar light, which is shown with the green contours. The X-ray emission appears to be symmetric, suggesting that the gas may be in hydrostatic equilibrium in the gravitational potential. The figure was adapted from \citet{bogdan13}.}
	\label{fig:ngc1961_ngc6753_images}
\end{figure}

\subsubsection{Massive disk galaxies}
\label{sec:overview_disk}

Although the ubiquity and luminous nature of the CGM around massive elliptical galaxies have been established based on observations with the \textit{Einstein} and \textit{ROSAT} X-ray satellites decades ago (see Section \ref{sec:overview_elliptical}), observations of disk-dominated galaxies were much less successful. However, these detections are especially important because the characterization of the CGM around massive disk-dominated galaxies can provide a unique insight into the formation and evolution of galaxies. 

The main advantage of disk-dominated galaxies is their environment. Because elliptical galaxies form through a series of mergers, they tend to reside in galaxy groups or galaxy clusters. Indeed, due the high galaxy density of rich environments, the likelihood of galaxy-galaxy interactions is much higher, thereby facilitating the formation of ellipticals. As opposed to this, disk-dominated galaxies may not undergo many merger events, hence a substantial population of these galaxies will reside in relatively isolated environments. In fact, the existence of a disk signifies that a galaxy did not experience strong mergers, because galaxy disks are destroyed during such violent events. Moreover, the relatively quiescent merger history of disk-dominated galaxies also assures that the characteristics of the CGM were not drastically altered by the energetic events that are typically initiated by mergers. 

A potential disadvantage of disk galaxies is the presence of other luminous X-ray emitting components, such as ultra-luminous X-ray sources or starburst-driven winds, that may either play an important role or even dominate the large-scale X-ray emission.  However, using various observational techniques, it is fairly easy to identify disk galaxies, whose overall emission is dominated by these X-ray emitting components. Excluding these -- typically very actively star-forming -- galaxies can result in a clean galaxy sample containing relatively quiescent isolated disk galaxies that should host quasi-static X-ray halos. 

While the bulk of the CGM is expected to originate from the accreted pristine gas at the earliest epochs of galaxy formation, the physical properties and spatial structure of the gas will reflect an imprint of the physical processes that shaped these galaxies. Thus, isolated disk-dominated galaxies are a unique population of galaxies that offer a view into the complex physical processes that influence the evolution of galaxies. 

Similar to ellipticals, disk-dominated galaxies were the subject of multiple observing campaigns carried out with all major X-ray telescopes. However, these searches were not conclusive and due to the lack of significant detections, serious doubts were cast on galaxy formation models and the basic principles of galaxy formation were questioned. Initial studies of disk-dominated galaxies used the \textit{ROSAT} X-ray observatory but failed to detect luminous X-ray halos \citep{benson00}. Because the CGM around these galaxies was fainter than predicted by early galaxy formation models, the search continued using more sensitive X-ray telescopes, such as \textit{Chandra} and \textit{XMM-Newton}. However, X-ray observations of ``normal'' (i.e.\ non-starburst) disk galaxies led to controversial results and non-detections \citep{rasmussen09}. While these attempts remained unsuccessful, a fair number of disk galaxies were shown to host X-ray gas. However, these disk galaxies have high star-formation rates, and the morphology of the gas exhibits a bipolar morphology, suggesting that it is driven by starburst \citep{strickland04,tullmann06}. This suggests that the bulk of the X-ray emission associated with the CGM of these galaxies is of internal origin (i.e.\ gas lifted by supernovae) and does not originate from the infall of primordial gas.

\begin{figure}[!tbp]
	\centering
	\vspace{-1cm}
	\begin{minipage}[b]{0.98\textwidth}
		\includegraphics[width=0.55\textwidth]{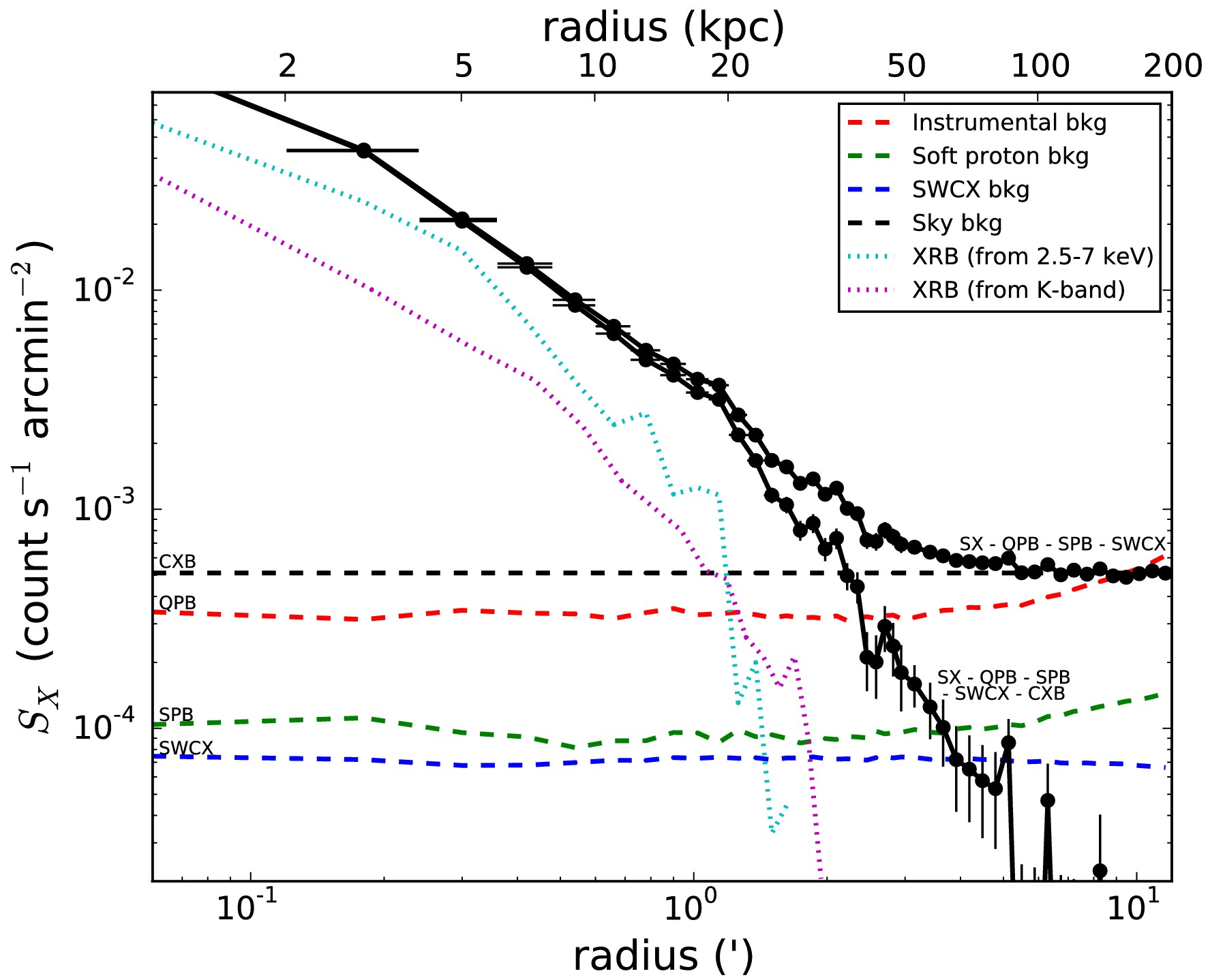}
	\end{minipage}
	\vspace{0cm}
	\caption{X-ray surface brightness profile  of the diffuse emission around NGC~1961 in the $0.4-1.25$~keV band based on \textit{XMM-Newton} data. The two sets of black data points show the background subtracted profiles. From the lower set of points all background components were subtracted. However, from the upper set of points the sky background was not removed. The diffuse emission can be robustly traced out to $\sim60$~kpc, beyond which radius the systematic uncertainties associated with the background subtraction start to dominate. The figure is adapted from \citet{anderson15}.}
	\label{fig:ngc1961_profile}
\end{figure}

The first success in the detection of the CGM around ``normal'' disk-dominated galaxies was achieved for NGC~1961 and NGC~6753 \citep{anderson11,bogdan13}. These galaxies are extremely massive ($M_{\rm \star} = (3-4)\times10^{11} \ \rm{M_{\odot}}$), fairly isolated, and have relatively low star formation rates ($\sim12-15 \ \rm{M_{\odot} \ yr^{-1}}$) given their stellar mass. Based on \textit{Chandra} and \textit{XMM-Newton} observations, the presence of luminous X-ray gas was demonstrated that exhibits a luminosity of $\sim6\times10^{40} \ \rm{erg \ s^{-1}}$ beyond the optical extent of the galaxies. The gaseous emission could be traced out to $\sim60$~kpc radius for both galaxies, implying that the gas is indeed extended and not confined to the optical body. However, it must be realized that this radius is still a relatively small fraction of the galaxy's virial radius ($\sim15\%r_{\rm 200}$). The azimuthal distribution of the CGM exhibits an approximately uniform distribution, suggesting that it resides in hydrostatic equilibrium rather than in a starburst-driven bipolar outflow. Thus, the detection of the large-scale CGM of these two massive galaxies confirmed that our basic picture of galaxy formation is correct and massive disk galaxies indeed host luminous X-ray emitting CGM.

Besides the detection of these X-ray halos, the physical characteristics of the X-ray gas could also be established, which were further improved using deep follow-up \textit{XMM-Newton} observations \citep{anderson15,bogdan17}. These studies established that the temperature of the gas is approximately consistent with the virial temperature of the dark matter halos and that the gas temperature slowly declines with increasing radius from $\sim0.7$~keV in the innermost regions to $\sim0.4$~keV at $\sim60$~kpc. Surprisingly, the metallicity of the gas was found to be strictly sub-Solar, $\sim(0.1-0.2)Z_{\rm \odot}$, at every radius, which is at odds with that obtained for massive elliptical galaxies, whose X-ray halos typically exhibit Solar metallicity. 
The physical properties of the gas and the implications of the results are further discussed in Section \ref{sec:metallicity}. In addition, based on the gas density and temperature profiles of the galaxies, their baryon mass fractions were estimated. \citet{bogdan17} found $f_{\rm b} \sim 0.08-0.1$, implying that NGC~1961 and NGC~6753 are missing about half of their baryons (see Section \ref{sec:baryons} for further discussion).   

These initial detections of the CGM were followed by other encouraging studies: extended hot gas was identified around UGC~12591 or NGC~266 \citep{dai12,bogdan13b}. A systematic study of six massive edge-on galaxies (including UGC~12591), the so-called Circum-Galactic Medium of MASsive Spirals (CGMMASS) sample, also demonstrated the presence of hot gas extending to $30-100$~kpc from the galactic center \citep{li17}. Although it was not possible to carry out in-depth studies of the CGM in these disk-dominated galaxies, the basic properties of the gas, such as the luminosity, temperature, gas density profile, could be established. Surprisingly, the X-ray luminosity of the hot gas around the CGMMASS galaxies was found to be much lower than that around NGC~1961 or NGC~6753. Overall, these detections hint that extended CGM is universal around massive disk-dominated galaxies. As a caveat, it must be noted that despite these successful detections, X-ray observations of several other (edge-on) disk galaxies did not result in statistically significant detections of the CGM \citep{bogdan15b}. This suggests that the highly luminous nature of the CGM around NGC~1961 and NGC~6753 may be the exception and not the rule and that the X-ray characteristics of the CGM exhibit notable galaxy-to-galaxy variations. Overall, despite the decades of endeavours, our understanding of the CGM around massive disk galaxies is far from being answered. 

Our in-depth knowledge about the large-scale CGM around massive disk galaxies is mostly based on the \textit{Chandra} and \textit{XMM-Newton} observations of NGC~1961 and NGC~6753 \citep{anderson13,bogdan17}. Interestingly, the properties of the CGM are markedly similar for both galaxies. The X-ray surface brightness profiles reveal the presence of diffuse emission beyond the stellar body ($\sim15$~kpc)  of the galaxies, which can be traced out to $\sim60$~kpc. While diffuse emission is likely to be present beyond this radius, it cannot be reliably mapped due to systematic uncertainties associated with the X-ray background subtraction. From the X-ray surface brightness profile, the assumed temperature, and metallicity profiles (see the next paragraphs), the gas density profile can be inferred. These measurements allowed to determine that the gas density drops from $n_{\rm e}\sim3\times10^{-3} \ \rm{cm^{-3}}$ to $n_{\rm e}\sim3\times10^{-4} \ \rm{cm^{-3}}$ between $10-60$~kpc. This measurement is of great importance for galaxy formation simulations: it carries crucial information about the energetic feedback from supernovae and AGN that shape the distribution of the hot gas. For example, more powerful feedback processes will push the gas to larger radii, resulting in more evacuated dark matter halos, hence lower densities and a more rapidly declining density profile. 

Due to the available deep X-ray observations, the temperature and metallicity profile of the CGM around NGC~1961 and NGC~6753 could also be measured. For both galaxies, the gas temperature profile exhibits a negative gradient, with the temperature decreasing from $kT\sim0.75$~keV in the innermost regions to $kT\sim0.4$~keV at about 50~kpc radius. This result is similar to that observed for elliptical galaxies that typically have flat or declining temperature profiles \citep{fukazawa06,humphrey06}.  

The deep X-ray observations of these galaxies also allow tracing the spatial distribution of the CGM. By studying the gas in circular surface brightness profiles, it was established that the X-ray gas has an approximately uniform distribution around the galaxies, which suggests that the gas is in hydrostatic equilibrium and is not dominated by a bipolar outflow driven by starburst or AGN. Similarly, the temperature structure of NGC~1961 and NGC~6753 was explored. Using simple one-component temperature models, it was established that the gas temperature exhibits some variations, which hint at the presence of a potentially more complex temperature structure. However, interpretation of these results is not trivial since the X-ray temperature measurements are luminosity-weighted, implying that the observed values are dominated by the gas at about $20-30$~kpc radius. To explore the temperature structure of CGM in more detail, a temperature map was computed for NGC~6753, which also highlighted the rather complex temperature structure of the CGM.

\section{Simulating X-ray halos around massive spiral galaxies}

Modern cosmological simulations include a wide range of physical processes ranging from supernova feedback to the feedback related to active galactic nuclei. Such hydrodynamic simulations have lead to impressive progress over the last decade producing galaxy populations that agree with many observable properties. These simulations can roughly be divided in two types: large volume and zoom-in simulations. The former aims to simulate large samples of galaxies simultaneously (e.g. IllustrisTNG, Eagle) whereas the latter focuses typically on individual galaxies (e.g. FIRE, Auriga) or galaxy clusters (C-Eagle). 

The simulations are based on three key ingredients: a cosmological framework, numerical discretization schemes for the matter components and the implementation of various astrophysical processes. The cosmological framework includes the nature of gravity (e.g. general relativity), the type of dark matter (e.g. cold dark matter) and the type of dark energy (e.g. a cosmological constant). Furthermore, the type of initial conditions also belong to the overall cosmological framework. The numerical discretization of the matter components (baryons, mostly in the form of gas, and dark matter) is typically implemented using a wide range of methods. The dark matter component is typically modelled with N-body methods representing a Monte Carlo scheme. The gas dynamics is modelled using different hydrodynamical methods. Most commonly employed methods are smoothed particle hydrodynamics (Lagrangian) and adaptive mesh refinement (Eulerian) methods. More recently also arbitrary Lagrangian-Eulerian methods have been employed, which use moving meshes for the underlying discretization. In addition to the hydrodynamics these models are also frequently extended to take into account the effects of magnetic fields, cosmic rays, radiation fields and cooling/heating processes. 

Most astrophysical processes are typically not resolved in cosmological simulations. They are therefore incorporated in the form of sub-resolution models. These processes include: star formation, stellar feedback, supermassive black holes, active galactic nuclei. Different simulations typically differ in how these sub-resolution models are constructed in detail. However, in all cases the models provide numerical closer at the resolution limit of the simulation, i.e. they represent the physics occurring on scales that cannot be resolved by the cosmological simulation.

Stellar feedback also plays an important role in regulating star formation, especially in lower mass systems. Also the implementation of stellar feedback requires sub-resolution models due to the numerical limitations of cosmological simulations. 
Any kind of interaction of stars with their surrounding gas contributes to the stellar feedback process. This occurs through the injection of energy and momentum that causes a regulation of the local star formation. Different numerical implementation have been developed to model the stellar feedback process in cosmological simulations.

\section{Confronting the observed and simulated properties of the CGM}

\subsection{X-ray scaling relations}
\label{sec:scaling}

Measuring the properties of the CGM around a statistically significant galaxy sample allowed to establish various X-ray scaling relations in elliptical galaxies (Section \ref{sec:overview_elliptical}). These relations connect the basic and easily observable properties of the hot gas, such as its luminosity or temperature, with the fundamental properties of galaxies, such as the stellar or virial mass. These scaling relations play an essential role in understanding various physical processes that influence the evolution of the hot X-ray emitting gas. By confronting the observed and simulated properties of the CGM, we can thoroughly probe simulations. Specifically, these test allow to verify the accuracy of a broad range of physics modules that play an essential role in shaping the hot gas content of galaxies. These processes include the heating from supernova and AGN feedback, enrichment by stellar and supernova yields, or mixing of the uplifted metals with the large-scale CGM.

The CGM around massive elliptical galaxies was routinely detected, which led to well-studied scaling relations of the hot X-ray gas (Section \ref{sec:overview_elliptical}). However, establishing scaling relations for large samples of disk-dominated galaxies is much more difficult, which is mostly due to the X-ray faint nature of these galaxies. However, it is important to realize that disk galaxies are not necessarily X-ray faint at the same stellar (or virial) mass, but they are -- on average -- less massive than elliptical galaxies, making them \textit{appear} fainter.

\begin{figure}[!tbp]
	\centering
	\vspace{-1cm}
	\begin{minipage}[b]{0.98\textwidth}
		\includegraphics[width=0.95\textwidth]{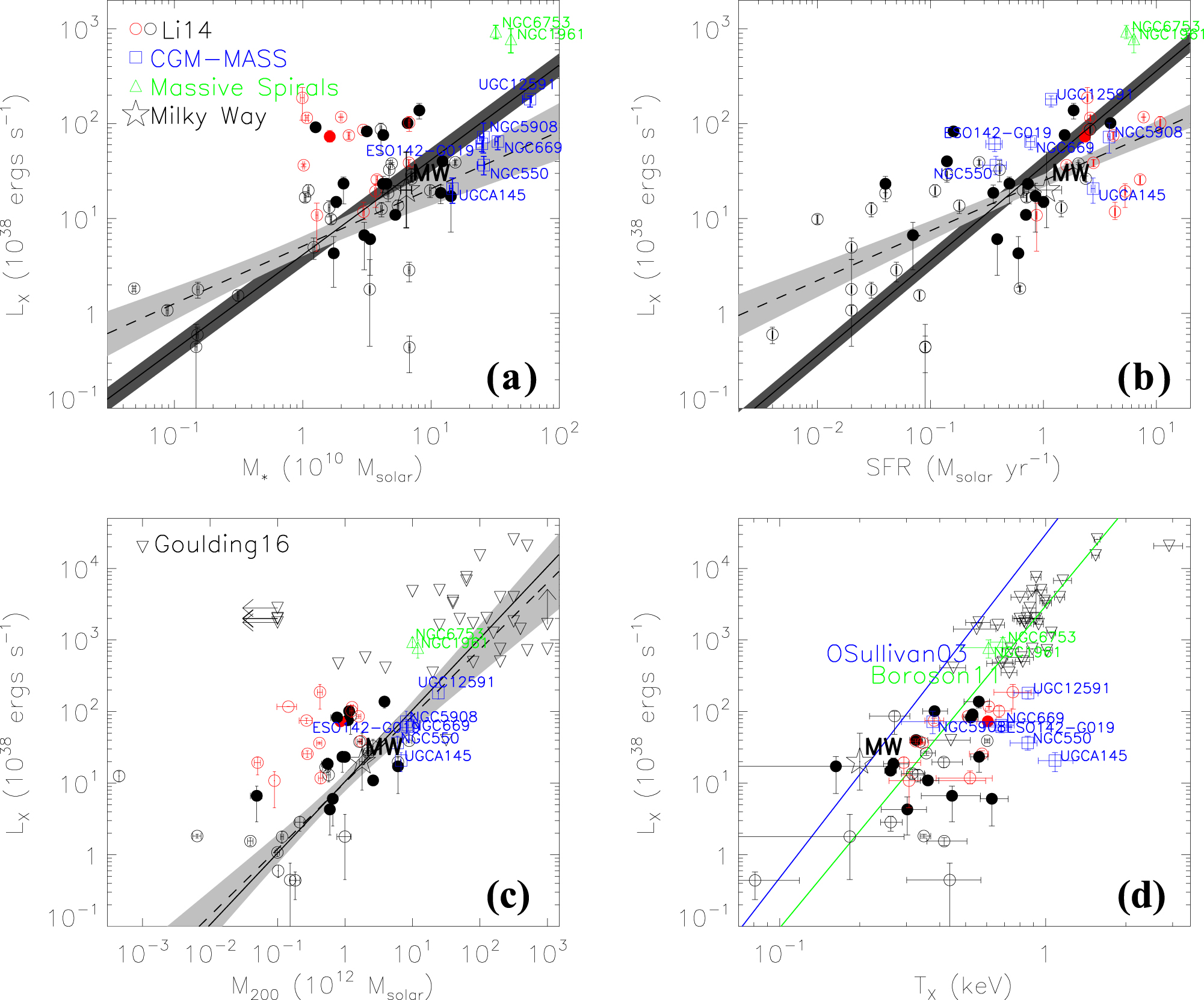}
	\end{minipage}
	\vspace{0cm}
	\caption{The $0.5-2$~keV band X-ray luminosity of the hot gas in the inner regions of galaxies ($<0.1r_{\rm 200}$) as a function of stellar mass (a) star-formation rate (b), virial mass (c), and gas temperature (d). Data points are taken for disk galaxies are taken from \citet{li14}, for the CGM-MASS galaxies from \citet{li17}, for massive spiral galaxies from \citet{bogdan13}, and for elliptical galaxies from \citet{goulding16}. The blue and green lines in panel (d) are relations from \citet{osullivan03} and \citet{boroson11}. The solid lines in panels (a)–(c) show the  best-fit relation to different sub-samples of \citet{li13}. The dashed line and shaded regions in panels (a)–(c) show the nonlinear fit to the data points and the $1\sigma$ confidence interval for the sub-samples. The plot was adapted from \citet{li17}.}
	\label{fig:scaling}
\end{figure}

Due to the faint nature of the CGM around spiral galaxies, the X-ray properties of the inner and outer halos were separately explored in many studies. A comprehensive study of the X-ray scaling relations of disk-dominated galaxies were carried out in \citet{li17}, who focused on the inner regions of a substantial sample of galaxies (Figure \ref{fig:scaling}). They included the X-ray emission from a wide range of galaxies, including highly inclined disk galaxies \citep{li14}, NGC~1961 and NGC~6753 \citep{anderson11,bogdan13}, CGM-MASS galaxies \citep{li17}, and, for comparison, the elliptical galaxies from the MASSIVE sample \citep{goulding16}. To keep the X-ray measurements consistent with each other, they restricted the comparison to the inner X-ray halo of the galaxies ($r < 0.1 r_{\rm 200}$). They concluded that the relation between the X-ray luminosity and stellar mass (or virial mass) exhibits a substantial scatter for disk galaxies, but they broadly follow the relation established for low-mass non-starburst galaxies. Interestingly, the most massive disk galaxies, NGC~1961 and NGC~6753, are outliers from these trends as they exhibit substantially higher X-ray luminosity. When compared with the population of elliptical galaxies from the MASSIVE sample, disk galaxies exhibit significantly lower X-ray luminosity, but they are also less massive than those studied by \citet{goulding16}, but overall follow the same $L_{\rm X} - M_{\rm 200}$ relation. When investigating the $L_{\rm X} - T_{\rm X}$ relation, several galaxies, such as the Milky Way, NGC~1961, NGC~6753, follow the relation established for elliptical galaxies. Surprisingly, most CGM-MASS galaxies in the sample of \citet{li17} exhibit hotter CGM than expected based on the relation with a temperature around $kT\sim0.8-1$~keV. 

\begin{figure}[!tbp]
	\centering
	\vspace{-1cm}
	\begin{minipage}[b]{0.98\textwidth}
		\includegraphics[width=0.6\textwidth]{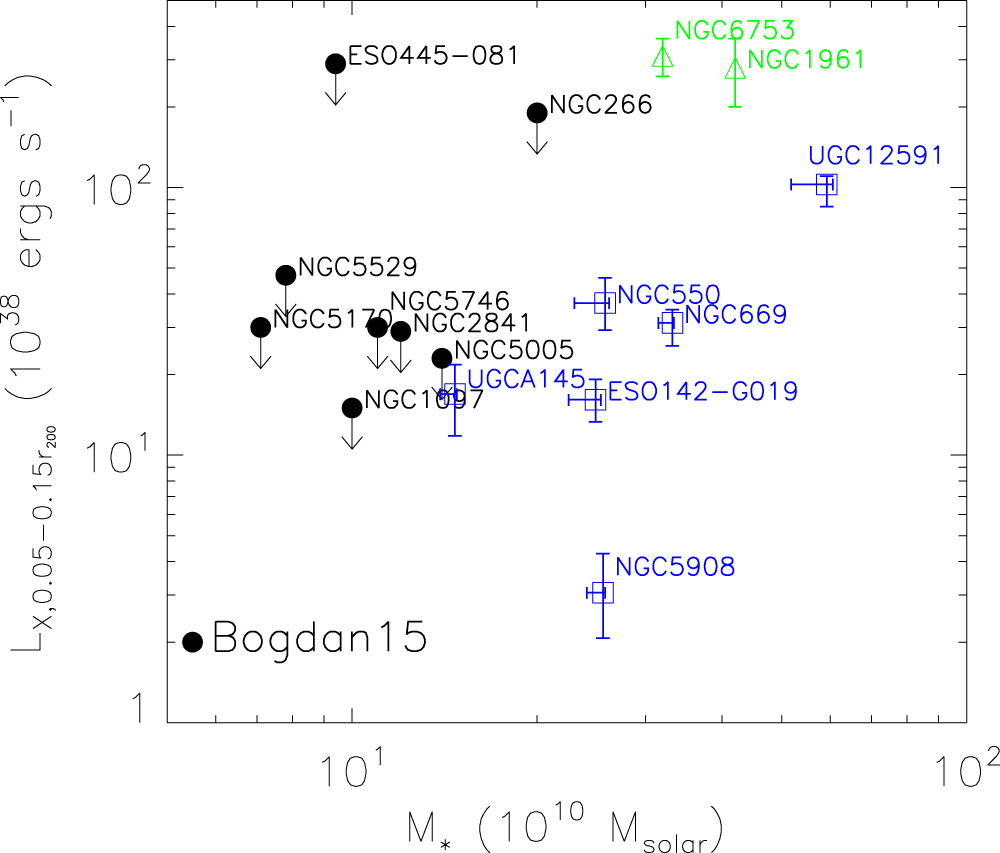}
	\end{minipage}
	\vspace{0cm}
	\caption{$0.5-2$~keV band X-ray luminosity of massive disk-dominated galaxies as a function of the stellar mass. The luminosity of the CGM was measured in the $r=(0.05-0.15)r_{\rm 200}$ region for all galaxies, which is beyond the stellar body of the galaxies. Most galaxies in the \citet{bogdan15b} sample (black data points) are not detected, except for NGC~1961 and NGC~6753 (green points). The data points for the CGM-MASS galaxies are taken from \citet{li17}. The figure is adapted from \citet{li17}.}
	\label{fig:lx_mstar_disk}
\end{figure}

Exploring the scaling relations in the outer halos ($0.05 r_{\rm 200}<r < 0.15 r_{\rm 200}$) around individual galaxies is much more complicated due to the very few detections (Figure \ref{fig:lx_mstar_disk}). For example, the $L_{\rm X} - M_{\rm \star}$ relation of detected galaxies only covers a narrow range, and galaxies with lower mass are not detected \citep{bogdan15b}. Interestingly, there is a large scatter in the X-ray luminosity of the handful of detected galaxies with the most significant outliers being NGC~1961 and NGC~6753. Therefore, it is not possible to identify statistically significant trends. However, all scaling relations suggest that the two most massive galaxies exhibiting the highest X-ray luminosity clearly do not represent the typical disk-dominated galaxies, but are over-luminous relative to the average population.

For galaxies with extended X-ray halos, the surface brightness profile can provide a wealth of information about the physical state of the gas. Most notably, the shape of the surface brightness profiles can inform simulations about the AGN feedback processes in disk galaxies. Therefore, it is interesting to probe whether the slope of the profiles exhibit significant variations, which may indicate differences in the AGN feedback. The slope of the surface brightness profile (i.e.\ the $\beta$ parameter) was compared for a sample of disk-dominated galaxies by \citet{li17}. They measured the best-fit parameter of $\beta=0.35-0.55$ for the disk galaxies in their sample, which is typical for galaxies. The obtained values did not reveal any correlation between the $\beta$ parameter and the physical properties (e.g.\ stellar mass) of the galaxies. As a caveat, we note that measurements of the $\beta$ parameter are affected by systematic uncertainties associated with the subtraction of the stellar and background components, which can significantly alter the observed values.

\subsection{Metallicity of the CGM}
\label{sec:metallicity}

Measuring the metallicity of the CGM can provide invaluable clues about the origin of the large-scale gas and can provide insights into the physicals processes that enrich the CGM and drive the metals to large radii. Therefore, probing the metallicity of the CGM around massive disk galaxies was in the major focus of observational studies. 

Due to the faint nature of the CGM at large radii, most of our knowledge comes from the deep \textit{XMM-Newton} observations of NGC~1961 and NGC~6753. These data allowed to study the X-ray spectrum of the hot gas in multiple regions, hence not only the average metallicity, but its distribution could also be measured. Surprisingly, both galaxies exhibit $Z\sim0.1-0.2 \ \rm{Z_{\odot}}$ abundance and the metallicity profile is flat at all radii. This result is in conflict with those obtained for massive elliptical galaxies, whose CGM typically exhibits Solar metallicity, but is more similar to X-ray faint ellipticals that have sub-Solar luminosity. The low metallicity around these massive galaxies suggests that the large-scale CGM is dominated by the primordial gas and it was not significantly enriched by metal-rich outflows driven by supernovae. 

As a caveat, we note that these measurements may be affected by the iron-bias, which appears if either a multi-temperature gas or a temperature gradient is described with a single component thermal model. For example, \citet{baldi06} investigated the Antennae galaxies by comparing results from the \textit{Advanced Satellite for Cosmology and Astrophysics} (\textit{ASCA}) and \textit{Chandra} and found that the complexity of the ISM can give low abundances when several different regions are averaged together. The multi-temperature nature of the gas is hinted by two observational facts. First, the large-scale CGM exhibits decreasing temperature with increasing radius (see the Section on Galaxy Clusters). Second, the innermost regions of the galaxies suggest that the ISM has two main components: one of them has high temperature and low abundance, while the other exhibits lower temperature and higher abundance \citep{anderson13}. In addition, the low observed metallicities may (partly) be explained by the incompleteness of atomic data tables for low temperature plasmas \citep{mernier20}. To further probe the metallicity of the gas, it would be ideal to utilize multi-temperature models in the outer regions of the galaxies, but due to the low signal-to-noise ratios of currently available observations, this is not feasible.  

The low metallicity of the CGM around these massive galaxies is rather surprising when we compare the total amount of metals in the CGM with that expelled by evolved stars and Type Ia supernovae (SN Ia). This conundrum can be best illustrated when estimating the amount of observed and expected iron in the CGM of NGC~1961 and NGC~6753. Assuming $0.1$~Solar abundance of the CGM and taking into account the abundance tables \citep{grevesse98}, we can derive the total iron mass within the virial radius, which is $\sim5\times10^7 \ \rm{M_{\odot}}$. Because the iron enrichment of the CGM occurs predominantly by SN Ia, we  can estimate the total iron yield by considering the SN Ia occurrence rate. The typical SN Ia rate per unit mass (SNuM) for S0a/b galaxies is $ 0.046^{+0.019}_{-0.017}$~SNuM, where 1 SNuM corresponds to  1 SN Ia per $10^{10} \ \rm{M_{\odot}}$ per 100 years \citep{mannucci05}. Considering the stellar mass of these galaxies, the SN Ia rate is about $0.02 \ \rm{yr^{-1}}$ (i.e.\ one SN Ia per 50 years). Because each SN Ia produces $\sim0.7 \ \rm{M_{\odot}}$ iron, the total iron mass in the galaxies can be produced within $\sim3$ Gyrs, which is much less than the age of the galaxies. In other words, based on the SN Ia occurrence rate, the expected metallicity of the gas should be significantly higher than the observed $\sim0.1 \ \rm{Z_{\rm \odot}}$.  

The discrepancy between the expected and observed iron abundance suggests that the bulk of the irons are either hidden from observations or are missing from the dark matter halos. The low iron abundances may be explained if the iron yields from SN Ia do not mix (well) with the hot gas, but instead cool to low temperatures \citep{brighenti05}. This, in turn, would render the iron ``invisible'' from X-ray observations and could result in the observed low metal abundances. If, however, the iron effectively mixes with the hot gas, but is not present in the CGM, it is likely that a notable fraction of the metals either left the galaxy or were expelled to larger radii in the form of a metal-rich outflow. In this case, a non-uniform metallicity profile would be expected with a positive metallicity gradient outwards (i.e.\ higher metallicity at large radii), which however is not observed either for NGC~1961 or NGC~6753. While higher metallicity may be present at radii $>60$~kpc, due to the low surface brightness of the CGM beyond $60$~kpc, present-day X-ray observatories cannot probe the presence and the metallicity at large radii. However, exploring this avenue will be the subject of future observations of the next generation of X-ray telescopes (see Section \ref{sec:future}). Because the observed metallicity profile is inconsistent with a pure metal-rich outflow, the low metallicities observed in the CGM of massive disk-dominated galaxies may be (partly) explained by metal-poor inflows of the pristine gas, which, in turn, should result in a negative metallicity gradient. While none of the above described scenarios can explain the observed low metallicities in itself, it is likely that the ecosystem of metals is much more complex, including inflows and outflows, mixing and stirring of the (primordial) hot gas with that expelled from the disk of the galaxies due to stellar and starburst driven winds and due to the energetic feedback from AGN.

Theoretically modelling metallicities in the CGM with hydrodyanmical simulations is a rather challenging problem. Incorporating the metal production in these models is in principle straightforward. However, significant uncertainties remain in terms of the exact metal yields and therefore the overall metal production of stars during their lifetime in the simulation. Furthermore, resolving the detailed phase structure of the gas in the CGM is also a numerical challenging for hydrodyanamical simulations. Therefore, refinement mechanisms are typically employed to increase the numerical resolution within the CGM \citep[e.g.]{vandevoort2019}.  

\begin{figure}[!tbp]
	\centering
	\vspace{-1cm}
	\begin{minipage}[b]{0.98\textwidth}
		\includegraphics[width=0.95\textwidth]{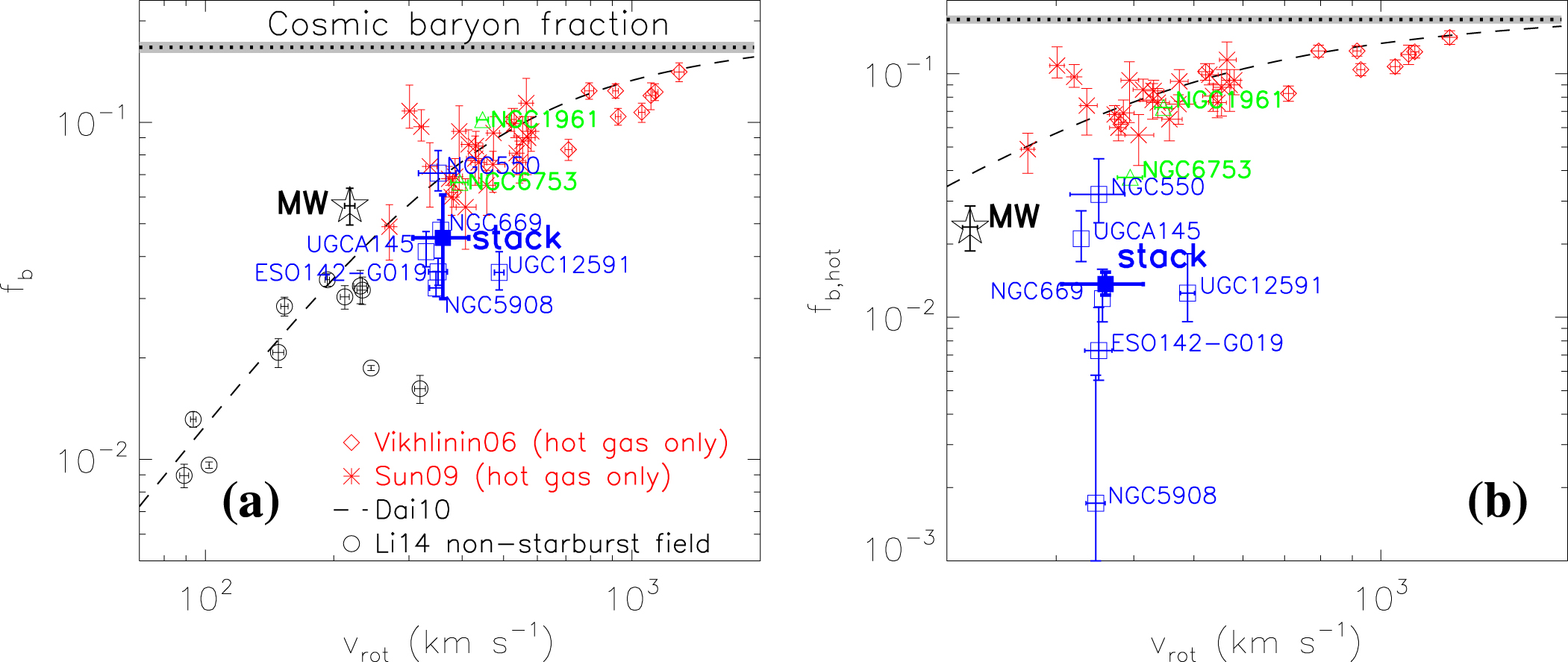}
	\end{minipage}
	\vspace{0cm}
	\caption{Baryon fraction of disk dominated galaxies as a function of the rotation velocity: for NGC~1961 and NGC~6753, for the CGM-MASS galaxies, and for low-mass non-starburst galaxies. The cosmic baryon mass fraction is shown with the dotted line with errors (shaded area). The left panel shows the baryon fraction of galaxies taking into account all components. Note that for galaxy groups and galaxy clusters, only the hot gas mass is included, which however dominates the baryon budget. The right panel shows the baryon fraction computed by considering only the hot gas mass for the disk galaxies. Note that massive disk dominated galaxies and even galaxy groups are missing fraction of their baryons. The plot was adapted from \citet{li18}.}
	\label{fig:baryon_fraction}
\end{figure}

\subsection{Missing baryon problem}
\label{sec:baryons}

In cosmology, we differentiate the missing baryon problems on different scales. The long-standing \textit{global} missing baryon problem is probably one of the most well-known challenges of modern astrophysics. Based on constraints from the Big Bang Nucleosynthesis \citep{fields14} and measurements from the cosmic microwave background \citep{planck16}, the baryon content of the high-redshift universe can be accurately determined. To maintain consistency with these constraints, we must be able to account for all the baryons even in the low-redshift ($z\lesssim2$) universe. However, when attempting to calculate the total baryon content at low redshifts, about one-third of the baryons appear to be missing, which defines the global missing baryon problem. 

Additionally, the \textit{local} missing baryon problem refers to a similar discrepancy, but on the scales of individual dark matter halos. According to the local missing baryon problem, the easily observable baryonic mass (stars, cool gas, hot gas in the centers of galaxies) of individual galaxies is much lower than expected based on the cosmic baryon fraction. This implies that either a substantial fraction of the baryons are hidden from observations or galaxies may have lost a fraction of the baryons throughout their evolution. In the following section, we focus on the local missing baryon problem and its potential resolution.

\subsubsection{Searching for the missing baryons with X-ray emission measurements}

Theoretical studies suggest that (at least fraction of) the baryons ``missing'' from the individual galaxies may reside in the form of tenuous hot gas. This gas is predicted to be very extended and it is believed to fill the entire dark matter halo of galaxies and may extend even beyond the virial radius of galaxies \citep{white91,fukugita06}. 

To derive the baryon mass fraction of galaxies, we must derive (1) the total gravitating mass of galaxies and (2) the total mass of all baryonic components. For disk-dominated galaxies, the total gravitating mass (or virial mass -- $M_{\rm 200}$) can be computed via multiple methods. The virial mass of disk galaxies could be derived from the baryonic Tully–Fisher relation for the cold dark matter cosmogony \citep{navarro97,mcgaugh00}, using the $M_{\rm 200} \propto V_{\rm max}^{3.23}$ relation, where $V_{\rm max}$ is the maximum
rotational velocity of the galaxies. The advantage of this method is that for disk galaxies it is relatively easy to determine the maximum rotational velocity using the rotation curves of galaxies. However, this method cannot be applied for elliptical galaxies, which systems are not supported by rotation. An alternative method to compute the virial mass of galaxies utilizes the hot X-ray gas around galaxies and makes the (reasonable) assumption that this gas is in hydrostatic equilibrium. If the gas is spherically symmetric, only the gas pressure is significant with other forces being insignificant, the mass profile of galaxies can be derived from the equation of hydrostatic equilibrium:  
\begin{equation}
    M(r) = \frac{k_{\rm B}T(r)r}{\mu m_{\rm p} G} \Big( \frac{d \log n_{\rm e}}{d \log r} + \frac{d \log T}{d \log r} \Big) \ ,
\end{equation}
where $\mu$ is the molecular weight, $T$ is the gas temperature, $k_{\rm B}$ is the Boltzmann constant, $\mu_{\rm p}$ is the proton mass, $n_{\rm e}$ is the electron density. Thus, the total mass of galaxies can be derived if the density and temperature of the X-ray gas are measured as a function of galactocentric radius. Based on X-ray observations, the derivative of the gas density is usually evaluated using a parameterized model (e.g.\ a $\beta$-model) for the gas density, where the parameters are determined by fitting the surface brightness profile of the X-ray gas. Similarly, the derivative of the temperature can be found using models that describe the temperature variation of the gaseous X-ray halo. The main advantage of this method is that it only employs X-ray measurements of the hot gas and is applicable to galaxies with all kinds of morphology. However, for low-mass galaxies, which lack a luminous X-ray halo, this method cannot be applied.  

To derive the baryon content of galaxies, all major baryonic components need to be accounted: the stellar mass of the galaxies, the cold gas, the hot X-ray emitting gas that may permeate the entire dark matter halo, and the mass of other less massive systems that reside within the galaxy's virial radius. We define the baryon mass fraction of galaxies as 
\begin{equation}
f_{\rm b} = M_{\rm b}/(M_{\rm b} + M_{\rm DM}) \ ,
\end{equation}
where $M_{\rm b}$ is the mass of baryons and $M_{\rm DM}$ is the mass of the dark matter and the sum of these two components is equivalent with the total gravitating mass. 

While it is relatively easy to account for the mass associated with stars and cold gas, precisely measuring the mass of the X-ray emitting hot gas presents a challenge. The main difficulty is due to the fact that X-ray emission studies only detect the gas out to a small fraction of the virial radius. However, the hot gas is expected to fill the entire dark matter halo and the bulk of the mass from this component will reside at large radii. To overcome this issue and derive the total hot gas mass, the density profiles can be extrapolated out to the virial radius, which can provide an estimate on the total mass. Using this approach, the baryon mass fraction of the massive disk-dominated galaxies, NGC~1961 and NGC~6753, were computed. For these galaxies a baryon mass fraction of ${f}_{{\rm{b}}}=0.05-0.08$ was obtained \citep{anderson15,bogdan17}. The sample of CGM-MASS galaxies also has similar, ${f}_{{\rm{b}}}\approx0.05$, baryon mass fractions \citep{li18}. Given that the cosmic value of the baryon mass fraction is ${f}_{{\rm{b}},{\text{}}\mathrm{WMAP}}=0.156\pm 0.002$ \citep{planck16}, we can conclude that massive disk galaxies are missing about half of their baryons and that the hot gas component cannot account for the missing baryons. In Figure \ref{fig:baryon_fraction} we present the baryon mass fraction obtained for a sample of disk-dominated galaxies, galaxy groups, and galaxy clusters.
 
Using a similar approach, \citet{humphrey06} computed baryon mass fractions for a sample of nearby massive elliptical galaxies. The virial mass of the galaxies was determined using the X-ray emitting gas and assuming that it is in hydrostatic equilibrium. After accounting for the baryonic components (most importantly the stars and the hot gas), they computed the baryon mass fraction and found that these galaxies have $f_{\rm b} = 0.04-0.09$. Thus, the results are good agreement with that obtained for disk-dominated galaxies. This suggests that galaxies of all morphological type are missing about half of their baryons. 

When considering the relative contribution of various baryonic components, it was found that the hot gas plays a major role. Specifically, in massive galaxies, the hot X-ray emitting gas can account for about half of the total baryons. However, as discussed above, the bulk of this gas cannot be directly observed by X-ray emission studies and its contribution is estimated by extrapolating the gas density and metallicity profiles out to the virial radius. Therefore, it is essential to briefly overview the main sources of uncertainties in the estimate of the total hot gas mass. The main source of uncertainty is associated with the unexplored nature of the hot gas beyond $\sim60$~kpc radius. Clearly, a significant temperature or metallicity gradient could result in different density profiles, and, hence total gas masses. Deviations from the hydrostatic equilibrium in the outskirts of the dark matter halos could also alter the total gas mass measurements. Finally, the true virial radius of galaxies may be different than the virial radius inferred from the the virial mass of the galaxies. A smaller/larger virial radius will result in a lower/higher total gas mass due to the different volume. 

Additionally, the measurement of the total gravitating mass from the hydrostatic equilibrium may also have substantial uncertainties. The main assumption of this calculation is that the X-ray emitting gas resides in hydrostatic equilibrium, which is the only (or dominant) source of pressure. Because the pressure may originate from other sources, such as magnetic fields, cosmic rays, or the bulk motion of the gas, the masses obtained through the simple assumption of hydrostatic equilibrium could be underestimated. To verify the accuracy of the masses obtained through this method, masses computed via velocity dispersion and weak lensing were compared with the X-ray estimates and it was found that they agree within a few tens of percent. The only significant differences were obtained for systems that undergo mergers, in which objects the assumption of hydrostatic equilibrium is clearly not valid. In general, simulations suggest that due to the bulk motions of the gas, the hydrostatic mass approximation will underestimate the mass by $5-20$\% \citep{nagai07}. However, this difference cannot account for the fact that massive  galaxies appear to be missing a substantial fraction of their baryons. 

Despite the above discussed uncertainties, the conclusion that individual galaxies are missing about half of their baryons is not entirely surprising. Indeed, a similar conclusion was obtained for galaxy groups, which also exhibit a baryon deficiency \citep[e.g.][]{giodini09}. The fact that massive galaxies not baryonically closed systems indicate that strong feedback from supernovae and AGN played a crucial role in the evolution of galaxies and pushed about half of the baryons beyond the virial radius of galaxies. 

Hydrodynamical cosmological simulations have proven to be a useful tool to investigate the location and distribution of baryons in the cosmic web \citep[e.g.][]{cen1999,dave2001,cen2006, shull2012,suresh2017}. In fact, several analyses based on hydrodynamical simulations predict that the missing baryons are in the form of warm-hot diffuse gas located between galaxies in the CGM and beyond the virial radius in the IGM. Such simulations have, for example, shown that star formation regulates
the production rate of metals, while processes such as supernova feedback, galactic winds, and AGN feedback are capable of expelling the enriched hot
gas into the CGM and IGM \citep[e.g.][]{cen2006,theuns2002,oppenheimer2006,wijers2020}.

\subsubsection{Searching for the missing baryons with X-ray absorption studies}

Because X-ray emission measurement of massive galaxies suggest that these systems are not barionically closed, the missing baryons were likely expelled from the galaxies and reside either around the dark matter halos of the galaxies or in the form of the warm-hot intergalactic medium (WHIM). Since the emission measure of the X-ray gas is proportional with the density square ($EM \propto n^2$), X-ray emission studies are less effective to probe the low-density gas around X-ray halos. A powerful method to probe the low-density gas in the outskirts of dark matter halos and beyond the virial radius of galaxies is to use X-ray absorption studies (see the chapter by Mathur in this Section). 

The X-ray halos of galaxies can imprint absorption lines on the X-ray spectrum of background quasars. Absorption studies in the UV wavelength, in particular with Hubble's Cosmic Origin Spectrograph, demonstrated the robustness of this approach. However, due to the much lower collecting area and spectral energy resolution of  X-ray grating instruments, detecting X-ray absorption lines from the CGM of galaxies has been a major challenge. A major success in this field was the detection of O~VII absorption lines from the hot gas around the Milky Way toward various AGN sightlines \citep{gupta12,gupta17}, which allowed to estimate the CGM mass around our own Galaxy . However, exploration of the CGM around external galaxies in absorption were much less successful. For example, \citet{yao10} probed the CGM along multiple luminous AGN sightlines, but could not obtain a statistically significant detection. In the absence of detections, they placed upper limits on the mean column densities of various ion species for the galaxies, such as the O~VII column density is $N_{\rm OVII} \leq 6\times10^{14} \ \rm{cm^{-2}}$. Converting this to an upper limit on the total mass, they concluded that the total mass enclosed in the CGM is 
\begin{equation}
M_{\rm gas} \lesssim 6\times10^{10} \ \rm{M_{\odot}} \Big(\frac{0.5}{f_{\rm O~VII}} \Big) \Big(\frac{0.3Z_{\odot}}{Z} \Big) \Big(\frac{R}{500 \ \rm{kpc}} \Big)^2  \ ,
\end{equation}
where $f_{\rm O~VII}$ is the ionization fraction of O~VII, $Z$ is the metallicity, and $R$ is the radius within which the mass of the CGM is measured. Using these characteristic values, we can conclude that it is unlikely that the bulk of the missing baryons reside in the outskirts of dark matter halos in low-density gas with temperatures of $10^{5.5}-10^{6.5}$~K, which is the sensitivity range of X-ray absorption studies. Thus, it is likely that a substantial fraction of baryons left the dark matter halos of galaxies. 

Using the \textit{Chandra} LETG observations of the luminous quasar, H1821+643, and a stacking approach, \citet{kovacs19} detected a $3.3\sigma$ detection of an O~VII absorption line originating from the outskirts of galaxies and the WHIM. The obtained column density is $N_{\rm OVII} = (1.4\pm0.4)\times10^{15} \ \rm{cm^{-2}}$, which exceeds the upper limit of \citet{yao10}. However, this signal is likely dominated by large-scale WHIM filaments rather than from the CGM of individual galaxies, which contributions cannot be easily differentiated. To further improve the constraints on the CGM in the outskirts of external galaxies, further absorption studies with the next-generation X-ray telescopes will be essential (Section \ref{sec:future}; also Chapter by Mathur, this Section).

\subsubsection{Sunyaev–Zel'dovich effect}

An alternative and powerful method to probe the low-density gas in the outskirts of dark matter halos and even beyond the galaxy's virial radius is to utilize the  Sunyaev–Zel'dovich effect. When photons from the cosmic microwave background pass through galaxies, galaxy groups, or galaxy clusters, they will be scattered by the free electrons in these objects. The thermal motion of the electrons will cause the so-called thermal SZE (tSZE) and the bulk motion of electrons will cause the kinetic SZE (kSZE) on the CMB photons. Therefore, probing the SZE of galaxies, galaxy groups, and galaxy clusters in the CMB allows us to probe the low-density gas associated with the extended dark matter halos of these galaxies.

The observed SZE signal ($Y$) for a single object, is proportional to the product of the total gas mass ($M_{\rm gas}$) and the virial temperature ($T_{\rm vir}$) and if $M_{\rm gas} \propto M_{\rm halo}$, we find that the SZE signal can be described as $Y\propto M_{\rm halo}^{5/3}$. For this reason, the most massive halos will have the highest signal, which in turn, makes them easily detectable. Indeed, SZE signal from individual galaxy clusters are routinely detected \citep{planck16}. 

Although the SZE signal associated with individual galaxies is much weaker and may remain below the detection threshold, the signal-to-noise ratios can be boosted by co-adding (i.e.\ stacking) the signal from a large number of individual galaxies. By performing a stacking analysis of a 260,000 locally brightest galaxies, including galaxy groups and individual galaxies, the SZE signal was detected for systems with masses as low as $M_{\rm 500} = 4\times 10^{12} \ \rm{M_{\odot}}$ \citep{planck13}. This mass is approximately comparable with the massive disk and elliptical galaxies observed in X-ray emission studies. Similar studies were performed by \citet{greco15}, who stacked the SZE signal of locally brightest galaxies to probe the hot gas around galaxies with similar halo masses. They also detected signals for galaxies with similar dark matter halo mass, but could not detect lower mass systems. When computing the total mass of the hot gas based on the SZE signal, it was established that the baryons missing from individual halos are, in fact, at larger scales.

While these studies stacked large samples of galaxies, the tSZE of nearby individual galaxies was also studied using Planck data. By analyzing a sample of 12 nearby spiral galaxies \citet{bregman21} detected SZE signal in the stack of these galaxies and could estimate the total gas mass within 250~kpc. They found that the total gas mass is about one-third of the predicted baryon content of the average galaxy in their sample. Based on their results, \citet{bregman21} suggests that the remaining baryons reside at an even larger radius and extend to the $400-500$~kpc volume. This conclusion is in agreement with that established based on large galaxy samples, and suggest that the a substantial fraction of the gas was indeed expelled beyond the virial radius of galaxies.

\begin{figure}[!tbp]
	\centering
	\vspace{-1cm}
	\begin{minipage}[b]{0.98\textwidth}
		\includegraphics[width=0.95\textwidth]{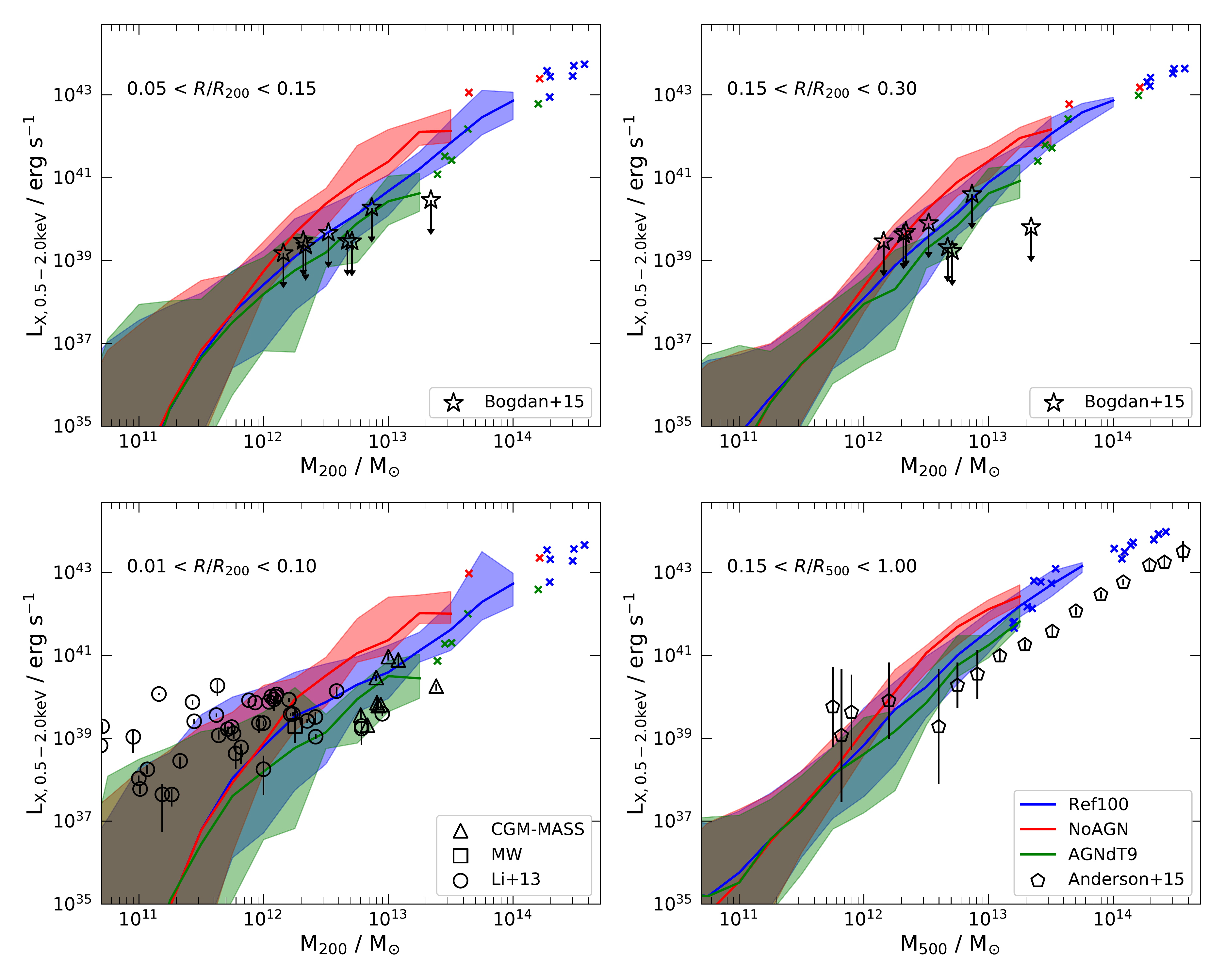}
	\end{minipage}
	\vspace{0cm}
	\caption{Comparison between the observed and predicted $0.5-2$~keV band X-ray luminosity of disk galaxies for different radial ranges. The specific regions were chosen to match the observational results from \citet{bogdan15b} (top panels), \citet{li17} (bottom left), and \citet{anderson15} (bottom right). The observed data points are shown with black symbols. The shaded regions show the $15$th and $85$th percentiles for simulated galaxies using three different feedback implementations: blue curves show the reference simulation, red curves show a simulation without AGN feedback, and the green curves show a modified AGN feedback in which each feedback event leads to a temperature change of $\Delta T_{\rm AGN} = 10^9$ K. The different prediction underscore the importance of feedback implementation. The figure was adapted from the work of \citet{kelly21}.}
	\label{fig:eagle_lx_m200}
\end{figure}

Hydrodynamical cosmological simulations have been used to study both the tSZE and kSZE.
For the tSZE signal from halos, the observational results and the predictions from simulations
indicate a certain degree of deviation from the self-similar case due to feedback mechanisms that strongly impact the gas content of halos with $M_{500}\sim 10^{12}$ \citep{lim21}.
Furthermore different simulations lead to  different tSZE signal predictions because of the
different implementation of the underlying physical model.
The Illustris simulation, for example, predicts a significantly lower thermal
energy of gas in halos with $M_{\rm 500} \sim 10^{13 - 13.5}$. This is
due to the stronger AGN feedback adopted in
that simulation compared to more recent simulations like IllustrisTNG. The predictions for $\tilde Y_500$ from IllustrisTNG and EAGLE on the other hand are remarkably similar. SZ observations can therefore also be used to constrain the underlying galaxy formation model. 

\subsection{The importance of AGN feedback on the observed properties of the CGM}

The energetic feedback from AGN plays a crucial role in many aspects of galaxy evolution from the earliest epochs to the present day. Naturally, these energetic events also have a profound effect on the observed properties of the CGM. Most importantly, the feedback energy from AGN can change the thermodynamic properties of the hot gas and it can drastically alter the distribution of the hot gas. The importance of energetic feedback from AGN was recognized based on different observational results. Here, we briefly discuss two of these observational evidences: the lack of intense star-formation associated with brightest cluster galaxies; and the surprisingly low X-ray luminosity of the hot CGM around individual galaxies. 

According to the well-known cooling flow problem, the central density of the intracluster medium (ICM) is low enough that it should cool on a timescale much shorter than the Hubble time, resulting in very rapid cooling onto the central central galaxy, the so-called brightest cluster galaxy, of the galaxy cluster. Considering the total mass of the ICM in the central regions of galaxy clusters, the average gas density, and taking into account the cooling time, the typical cooling rate of the gas should be $100–1000 \ \rm{M_{\odot} \ yr^{-1}}$. This, in turn, would suggest that the central regions of galaxy clusters exhibit strong cooling flows and BCGs should have extremely high star-formation rates \citep{fabian77,cowie77,fabian94}. However, a wide range of \textit{Chandra} and \textit{XMM-Newton} observations demonstrated that only a small fraction of gas cools to low temperatures and provides material to star-formation \citep{david01,peterson03}. To resolve the cooling flow problem, it was suggested that the hot gas in the central regions of galaxies are re-heated by the feedback energy from AGN \citep{churazov01,peterson06}. The observational signatures of AGN feedback are most apparent in the large-scale distribution of the hot gas on galaxy, galaxy groups, and galaxy cluster scales. Specifically, AGN inflates large bubbles in the X-ray gas, which rise buoyantly to a large radius \citep{jones02,david11,randall11,blanton11,fabian12}. In addition, the energy from AGN is also transported to the X-ray gas via shocks and turbulent mixing. According to theoretical calculations, it was established that the mechanical energy from AGN can offset the radiative losses due to cooling, implying that this energetic feedback can prevent the runaway cooling of the central gas \citep{mcnamara07,mcnamara12}. Since this energetic feedback is present in both galaxies, galaxy groups, and galaxy clusters, this underlines the importance of accurate modeling of the AGN feedback.

Although the first analytic description of galaxy formation models suggested that the CGM should be ubiquitous around individual galaxies, the predicted X-ray luminosity of these gaseous halos were overestimated. For example, \citet{white91} predicted that a dark matter halo with a circular velocity of $300 \ \rm{km \ s^{-1}}$ should have an X-ray luminosity of $3\times10^{42} \ \rm{erg \ s^{-1}}$. Such high x-ray luminosity around disk-dominated galaxies should have been detected by the \textit{ROSAT} X-ray observatory, but these galaxies remained undetected. The reason for the overly high predicted X-ray luminosity is that they neglected gas ejection and assumed that the hot gas follows the dark matter distribution. Hydrodynamical simulations performed by \citet{toft02} predicted about two orders of magnitude lower X-ray luminosity for the CGM component. However, the main reason of the low X-ray luminosity was the absence of energetic feedback, which, in turn, resulted in an overly massive stellar component and therefore a less massive and less luminous CGM component. The X-ray properties of disk-dominated galaxies were simulated by \citet{crain10}, who incorporated efficient feedback from  supernovae. This feedback prevents the conversion of halo gas into stars, hence prevents the overcooling and the formation of an overly massive stellar component. In addition, the entropy profile of the CGM is also changing due to the feedback energy, as a fraction of the baryons are expelled to larger radii resulting in shallower gas density profiles in the central regions of galaxies. However, these simulations did not include the energetic AGN feedback, which further alters the predicted properties of the CGM. Specifically, powerful AGN feedback is likely responsible for the fact that even the most massive galaxies appear to be missing a substantial fraction of their baryons. Indeed, due to the energy input from AGN, about half of the baryons were expelled beyond the virial radius of galaxies. 

Hydrodynamical simulations also incorporate AGN feedback due to supermassive black holes. 
However, numerically resolving the relevant scales is particularly problematic for AGN feedback. For example, the highly-collimated jets of relativistic particles can in general cosmological simulations not be resolved because jets themselves cover an enormous dynamic range, being launched at several Schwarzschild radii, and propagating outwards to tens of kpc. The feedback of active galactic nuclei is commonly divided in two models: quasar and radio. Quasar mode feedback represents the efficient mode of black hole growth and is typically implemented through energy or momentum injection. It is assumed that the bolometric luminosity is proportional to the accretion rate, which is in the simplest case derived from a Bondi-Hoyle accretion model. Some fraction of this energy is then deposited in the form of feedback energy. Highly collimated jets of relativistic particles lead radio mode feedback, which is often associated with X-ray bubbles that can offset cooling losses. This mode of feedback is particularly important to regulate star formation in massive galaxies.

The prescription to incorporate AGN feedback (or the complete lack thereof) in galaxy formation simulations, drastically alters the predicted physical properties of the CGM. The most simple observational tests to probe the accuracy of AGN feedback models are to (1) measure the X-ray luminosity of the CGM, and (2) probe the X-ray gas density (or entropy) profile. Both of these observables strongly depend on the implementation of the AGN feedback. For example, if the AGN feedback is weak (or completely lacking), galaxies will host a large amount of X-ray gas in their dark matter halo, resulting in high CGM luminosity. If, however, the AGN feedback is strong, the hot gas will be expelled to the outskirts of the dark matter halo or could be completely removed from the halo, which will result in a low X-ray luminosity. Similarly, the X-ray gas density profile will also depend on the feedback implementation: powerful feedback will result in shallower profiles in the central regions of the galaxies.    

Based on these considerations, the characteristics of the CGM around disk-dominated galaxies were compared with modern galaxy formation simulations. For example, the X-ray properties of a substantial sample of disk galaxies were compared with the Galaxies–Intergalactic Medium Interaction Calculation (GIMIC) simulation \citep{li14}. While there was a reasonable agreement between the observed and predicted X-ray properties, the scatter in the $L_{\rm{X}} - M_{\rm 200}$ $L_{\rm{X}} - SFR$ was extremely small and was inconsistent with observations. This pointed out a major shortcoming of the GIMIC simulation, namely the lack of AGN feedback and the adoption of constant stellar feedback parameters. In a similar study, \citet{bogdan15b} utilized a sample of massive disk-dominated galaxies to probe the Illustris simulation. They found that the simulation broadly agrees with the observed X-ray luminosity and upper limits of the disk galaxies. However, for the most massive galaxies, the predicted X-ray luminosity of the CGM fell short of the observed values. This suggested that the AGN in these massive galaxies provided overly powerful radio-mode feedback, which pushed the gas out from the dark matter halo, and, hence resulted in low X-ray luminosity. These comparisons illustrate the powerful constraining nature of observations of the CGM. Indeed, the comparison between observational studies and theoretical predictions played an important role in revising these simulations, and essentially in the production of the EAGLE and IllustrisTNG simulation suites.

\begin{figure}[!tbp]
	\centering
	\vspace{-1cm}
	\begin{minipage}[b]{0.98\textwidth}
		\includegraphics[width=0.95\textwidth]{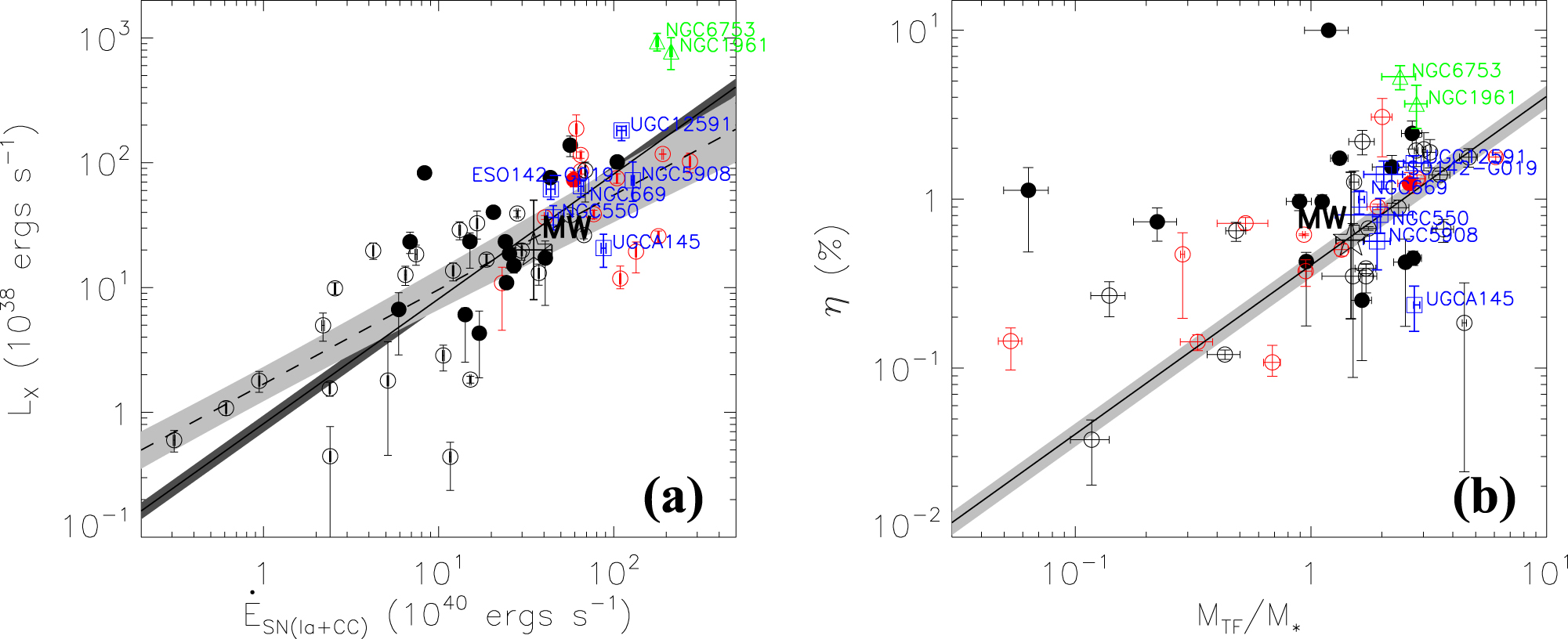}
	\end{minipage}
	\vspace{0cm}
	\caption{Energy budget of disk galaxies. The left panel shows the X-ray luminosity of disk galaxies measured within $r<0.1r_{\rm 200}$ as a function of the total energy injection from supernovae (core collapse and Type Ia). The right panel shows the X-ray coupling efficiency as a function of baryonic to stellar mass ratio of the galaxies. The dashed line and shaded regions represent the nonlinear fit to the data points and the $1\sigma$ confidence interval for the sub-samples in the \citet{li13} sample.  The figure was adapted from \citet{li17}.}
	\label{fig:energy_budget}
\end{figure}

\subsection{Missing feedback problem}

While the large-scale CGM of disk galaxies is in a quasi-static state, the inner regions of disks are affected by very complex physical processes. Notably, there is an active interplay between the CGM, the stellar ejecta from evolved stars and supernovae, and the energy input from supernova feedback, which are capable to drive galactic-scale winds. When considering the energy input from supernovae and comparing it with the X-ray luminosity of the hot gas in the innermost regions of disk galaxies, a stark discrepancy is observed. Specifically, the coupling efficiency, which is defined as the ratio of the X-ray luminosity of the hot gas to the energy input from supernovae, is only $\sim0.004$ \citep{li13}. This surprisingly low value implies that the bulk of the supernova energy is missing, which is often referred to as the ``missing feedback'' problem. 

To understand how the missing feedback problem could be resolved, we must consider how the supernova-heated gas may evolve. First, the hot gas could be expelled from the dark matter halo of the galaxy and become part of the large-scale galaxy group or cluster atmosphere or it could join the warm-hot intergalactic medium. In this case, the hot gas will reside beyond the galaxy's virial radius, and, hence it will be ``hidden'' from the X-ray observations of galaxies. Second, the gas could be ejected from the galactic disk, but stay associated with the dark matter halo of the galaxy. In this scenario, the metal-enriched uplifted gas will mix with the CGM that was previously dominated by the infalling pristine gas. Third, the outflowing hot gas could rapidly cool down in the proximity of the disk and then fall back to the galaxy disk, thereby re-joining the interstellar medium. 

To differentiate between these scenarios, the characteristics of the hot X-ray gas (both in the disk and on larger scales) must be investigated. For example, if the gas is ejected to large radii or even expelled from the dark matter halo, the amount and the luminosity of the X-ray emitting gas in the innermost regions will decrease. Additionally, if the gas leaves the central regions of the galaxy, the density profile of the gas should be flatter in the innermost regions should further decrease outwards. If, however, the gas falls back to the disk in a galactic fountain, high density gas and luminous X-ray emission is expected in the central regions of the galaxy. Considerations of the energy budget can also hint at the fate of the supernova heated gas. Because of the gravitational potential of galaxies and the available feedback energy from supernovae, the gas from the innermost regions can only leave the dark matter halos of low-mass galaxies but cannot leave the dark matter halos of massive galaxies  \citep{li17}. In addition, because of the inefficient nature of gas cooling, the rate of infall to the disk is also low, which makes the galactic fountain scenario unlikely. Taken these together and considering the observed X-ray properties of disk galaxies, the most plausible scenario is that the gas heated by the energetic feedback of supernovae remains in the dark matter halo of massive galaxies and mixes with the CGM. 

The coupling efficiency was measured for a large sample of disk galaxies by \citet{li13,li17} (Figure \ref{fig:energy_budget}). They found that this parameter exhibits an increasing trend with the total mass of galaxies \citep{li13}. This can be attributed to the deeper potential well of more massive galaxies: they can retain more gas heated by the supernova feedback. Hydrodynamical simulations can directly probe the X-ray coupling efficiency as a function of the total mass of the galaxies. Based on the EAGLE simulation, \citet{kelly21} found that the coupling efficiency -- in agreement with the observations -- increases with the total mass of galaxies. Upon investigating the evolution of the gas due to the energetic feedback processes, they concluded that the bulk of the feedback energy is lost as hot gas is ejected from the galaxy disks. In the case of low mass dark matter halos ($M \lesssim 10^{11} \ \rm{M_{\odot}}$), the gas can be completely ejected from the halo on a timescale of $\sim10^8$ years. However, for more massive galaxies, the supernova feedback energy is not sufficient to drive the gas from the dark matter halo. This conclusion is in agreement with the energetic considerations discussed above \citep{li17}. In these massive galaxies, the gas will be retained in higher density regions, where it can radiate the energy injected by supernovae, which in turn results in higher X-ray luminosity. This theoretical picture is consistent with the fact that the most massive disk-dominated galaxies, NGC~1961 and NGC~6753, have much higher X-ray luminosity than their lower-mass counterparts.

\section{Future outlook}
\label{sec:future}

The X-ray halos of galaxies were studied in the past with every major X-ray observatory, including \textit{ROSAT}, \textit{Chandra}, and \textit{XMM-Newton}. All these telescopes 
had various advantages that contributed to the understanding of the X-ray halos. Notably, \textit{ROSAT} allowed the detection of large samples of galaxies and could provide an volume-limited galaxy sample that was free from selection effects. However, due to the low sensitivity and poor spatial resolution, observations taken by \textit{ROSAT} could not collect a large number of photons from the CGM of galaxies and it was not possible to identify contaminating sources in the emission associated with galaxies. Overall, these complications made it challenging to interpret the results obtained through the \textit{ROSAT} All-Sky Survey. Due to the superior, $0.5''$, angular resolution of \textit{Chandra}, it became possible to detect bright point sources, notably AGN, LMXBs, and HMXBs, associated with galaxies and exclude them from the study of the diffuse emission. Because of the understanding of the X-ray scaling relations of X-ray binaries with the stellar mass and star-formation rate, it even became possible to estimate the contribution of unresolved X-ray binaries to the overall emission level. This, in turn, allowed to precisely account for the population of resolved and unresolved X-ray sources and resulted in the accurate study of the truly diffuse emission. However, the identification of discrete sources required deep observations, implying that it was difficult to study large galaxy samples. In addition, many studies focused only on the most nearby and brightest galaxies, hence the galaxy sample observed by \textit{Chandra} is hampered by a wide range of selection effects. While there were attempts to carry out more systematic surveys, such as the ${\mathrm{ATLAS}}^{{\rm{3D}}}$ or MASSIVE surveys, these still did not provide a large galaxy sample covering galaxies with various stellar masses and star-formation rates. \textit{XMM-Newton} with its large collecting area allowed the detection of relatively faint emission from X-ray halos in the outskirts of galaxies, as discussed for the massive spiral galaxies NGC~1961 or NGC~6753. However, the relatively high instrumental background of this telescope limits the sensitivity of the observations since the emission at the outskirts of galaxies becomes dominated by systematic uncertainties. With its modest angular resolution, it was possible to identify some of the point sources, but many of them remained unresolved and contributed to the truly diffuse emission. Similarly to \textit{Chandra} this pointing X-ray telescope also studied individual galaxies rather than carrying out large systematic surveys, implying that the \textit{XMM-Newton} archive cannot be used to study the X-ray halos of galaxies in an unbiased manner. 

The understanding of the X-ray halos will be dramatically improved by observations carried out by the Spektr-RG (SRG) telescope, which was launched in July 2019. The main instrument aboard SRG is the eROSITA telescope that consists of seven identical mirror modules, providing excellent sensitivity. During its first four years of operation, eROSITA will carry out 8 full-sky surveys - once every 6 months. Given the high sensitivity of eROSITA, this survey will be $30-50$ times more sensitive than its predecessor, the ROSAT All-Sky Survey. In addition to the greater sensitivity, eROSITA also has $\sim3$ times the higher angular resolution than ROSAT hence it will be possible to identify bright point sources. Thanks to the greater sensitivity and our much-improved understanding of the X-ray binary population, it will be possible to carry out an unbiased volume-limited study of the CGM around large galaxy samples in the nearby Universe. For example, it will be possible to probe how the X-ray emission of gaseous halos scales with the stellar mass, star-formation rate, or morphology based on large galaxy samples in a statistical manner -- similar to studies that were carried out in optical waveband using the wealth of SDSS data. For example, \citet{oppenheimer20} suggested based on state-of-the-art X-ray simulations that the X-ray halos of massive nearby galaxies will be observed out to $\sim150$~kpc radii when stacking the X-ray photons of eROSITA galaxies. 

On a longer timescale, the next-generation X-ray telescopes can completely revolutionize our understanding of the X-ray halos of galaxies. The proposed ARCUS X-ray observatory will study the CGM of galaxies in absorption: it will use bright background quasars that illuminate the hot gas in the outskirts of galaxies (and galaxy clusters). Thanks to the much larger collecting area of ARCUS and by studying multiple quasar sightlines, it will be able to map the structure, the temperature, and even the gas motions of the X-ray gas. This will not only allow to address the missing baryon problem on galaxy scales but can also lead to a much better understanding of the metal cycling in and out of galaxies. These studies will be further elevated by the proposed Lynx observatory that will have large throughput and sub-arcsecond angular resolution. Therefore, Lynx will be able to probe the X-ray halos of galaxies even around low-mass (Milky Way-type) galaxies close to their virial radius, which remains impossible with the present generation telescope. Such studies will be carried out both in emission and absorption using background quasars. This will allow the detailed mapping of energetic feedback on galaxy scales: in low-mass galaxies, winds driven by supernovae and in massive galaxies, the effects of supermassive black holes will be mapped. This will lead to a comprehensive picture of the most fundamental processes that influence the evolution of galaxies from their birth to the present day. 

On the theoretical frontier, hydrodynamical simulations have to incorporate more fine-grained and predictive models that are less calibration dependent. Currently, large-volume simulations can provide large samples of galaxies. However, these simulations rely on rather crude models that have to be calibrated against a few key observables. On the other hand, the galaxy formation models of high resolution simulations are less calibration dependent and allow more detailed modelling of the actual physical processes. Ideally, these two approaches should be combined in the near future to allow simulations of large samples of galaxies with detailed galaxy formation models.

\noindent
\begin{center}
%==================================================\\
%MARK IS WRITING HERE \\
%================================================== \\
\end{center}

%{\bf X-ray predictions of cosmological simulations}:
%
%
%

%
%
%

%\begin{itemize}
%    \item Early attempts to simulate galaxy halos. Big discrepancy between very large predicted luminosities and non-detected spiral galaxies, as simulations predicted very luminous X-ray halos. The importance of feedback in resolving the discrepancy: the gas is pushed out to large radii.
%    \item Description of modern simulations and their physics, notably AGN and SN feedback. IllustrisTNG and EAGLE are the most state-of-the-art simulation that contain more complete physics. Predictions on the X-ray luminosities  of luminous X-ray halos were substantially lowered due to the introduction of feedback.
%    \item Comparison of observations and simulations. We can compare luminosities, temperatures, metallicities, density profiles. The agreement is better with modern simulations than before, but there are still many challenges present. Feedback can be implemented in many different ways.
%\end{itemize}

\bibliographystyle{apj}
\bibliography{bibliography.bib} % your references Yourfile.bib

\begin{thebibliography}{}
\expandafter\ifx\csname natexlab\endcsname\relax\def\natexlab#1{#1}\fi

\bibitem[{{Anderson} \& {Bregman}(2011)}]{anderson11}
{Anderson}, M.~E., \& {Bregman}, J.~N. 2011, ApJ, 737, 22

\bibitem[{{Anderson} {et~al.}(2013){Anderson}, {Bregman}, \&
  {Dai}}]{anderson13}
{Anderson}, M.~E., {Bregman}, J.~N., \& {Dai}, X. 2013, ApJ, 762, 106

\bibitem[{{Anderson} {et~al.}(2016){Anderson}, {Churazov}, \&
  {Bregman}}]{anderson15}
{Anderson}, M.~E., {Churazov}, E., \& {Bregman}, J.~N. 2016, MNRAS, 455, 227

\bibitem[{{Baldi} {et~al.}(2006){Baldi}, {Raymond}, {Fabbiano}, {Zezas},
  {Rots}, {Schweizer}, {King}, \& {Ponman}}]{baldi06}
{Baldi}, A., {Raymond}, J.~C., {Fabbiano}, G., {et~al.} 2006, ApJS, 162, 113

\bibitem[{{Benson} {et~al.}(2000){Benson}, {Bower}, {Frenk}, \&
  {White}}]{benson00}
{Benson}, A.~J., {Bower}, R.~G., {Frenk}, C.~S., \& {White}, S.~D.~M. 2000,
  MNRAS, 314, 557

\bibitem[{{Blanton} {et~al.}(2011){Blanton}, {Randall}, {Clarke}, {Sarazin},
  {McNamara}, {Douglass}, \& {McDonald}}]{blanton11}
{Blanton}, E.~L., {Randall}, S.~W., {Clarke}, T.~E., {et~al.} 2011, ApJ, 737,
  99

\bibitem[{{Bogd{\'a}n} {et~al.}(2017){Bogd{\'a}n}, {Bourdin}, {Forman},
  {Kraft}, {Vogelsberger}, {Hernquist}, \& {Springel}}]{bogdan17}
{Bogd{\'a}n}, {\'A}., {Bourdin}, H., {Forman}, W.~R., {et~al.} 2017, ApJ, 850,
  98

\bibitem[{{Bogd{\'a}n} {et~al.}(2013{\natexlab{a}}){Bogd{\'a}n}, {Forman},
  {Kraft}, \& {Jones}}]{bogdan13b}
{Bogd{\'a}n}, {\'A}., {Forman}, W.~R., {Kraft}, R.~P., \& {Jones}, C.
  2013{\natexlab{a}}, ApJ, 772, 98

\bibitem[{{Bogd{\'a}n} \& {Goulding}(2015)}]{bogdan15a}
{Bogd{\'a}n}, {\'A}., \& {Goulding}, A.~D. 2015, ApJ, 800, 124

\bibitem[{{Bogd{\'a}n} {et~al.}(2013{\natexlab{b}}){Bogd{\'a}n}, {Forman},
  {Vogelsberger}, {Bourdin}, {Sijacki}, {Mazzotta}, {Kraft}, {Jones},
  {Gilfanov}, {Churazov}, \& {David}}]{bogdan13}
{Bogd{\'a}n}, {\'A}., {Forman}, W.~R., {Vogelsberger}, M., {et~al.}
  2013{\natexlab{b}}, ApJ, 772, 97

\bibitem[{{Bogd{\'a}n} {et~al.}(2015){Bogd{\'a}n}, {Vogelsberger}, {Kraft},
  {Hernquist}, {Gilfanov}, {Torrey}, {Churazov}, {Genel}, {Forman}, {Murray},
  {Vikhlinin}, {Jones}, \& {B{\"o}hringer}}]{bogdan15b}
{Bogd{\'a}n}, {\'A}., {Vogelsberger}, M., {Kraft}, R.~P., {et~al.} 2015, ApJ,
  804, 72

\bibitem[{{Boroson} {et~al.}(2011){Boroson}, {Kim}, \& {Fabbiano}}]{boroson11}
{Boroson}, B., {Kim}, D.-W., \& {Fabbiano}, G. 2011, ApJ, 729, 12

\bibitem[{{Bregman} {et~al.}(2021){Bregman}, {Hodges-Kluck}, {Qu}, {Pratt},
  {Li}, \& {Yun}}]{bregman21}
{Bregman}, J.~N., {Hodges-Kluck}, E., {Qu}, Z., {et~al.} 2021, arXiv e-prints,
  arXiv:2107.14281

\bibitem[{{Brighenti} \& {Mathews}(2005)}]{brighenti05}
{Brighenti}, F., \& {Mathews}, W.~G. 2005, ApJ, 630, 864

\bibitem[{{Cen} \& {Ostriker}(1999)}]{cen1999}
{Cen}, R., \& {Ostriker}, J.~P. 1999, ApJ, 514, 1

\bibitem[{{Cen} \& {Ostriker}(2006)}]{cen2006}
---. 2006, ApJ, 650, 560

\bibitem[{{Churazov} {et~al.}(2001){Churazov}, {Br{\"u}ggen}, {Kaiser},
  {B{\"o}hringer}, \& {Forman}}]{churazov01}
{Churazov}, E., {Br{\"u}ggen}, M., {Kaiser}, C.~R., {B{\"o}hringer}, H., \&
  {Forman}, W. 2001, ApJ, 554, 261

\bibitem[{{Cowie} \& {Binney}(1977)}]{cowie77}
{Cowie}, L.~L., \& {Binney}, J. 1977, ApJ, 215, 723

\bibitem[{{Crain} {et~al.}(2010){Crain}, {McCarthy}, {Frenk}, {Theuns}, \&
  {Schaye}}]{crain10}
{Crain}, R.~A., {McCarthy}, I.~G., {Frenk}, C.~S., {Theuns}, T., \& {Schaye},
  J. 2010, MNRAS, 407, 1403

\bibitem[{{Dai} {et~al.}(2012){Dai}, {Anderson}, {Bregman}, \&
  {Miller}}]{dai12}
{Dai}, X., {Anderson}, M.~E., {Bregman}, J.~N., \& {Miller}, J.~M. 2012, ApJ,
  755, 107

\bibitem[{{Dav{\'e}} {et~al.}(2001){Dav{\'e}}, {Cen}, {Ostriker}, {Bryan},
  {Hernquist}, {Katz}, {Weinberg}, {Norman}, \& {O'Shea}}]{dave2001}
{Dav{\'e}}, R., {Cen}, R., {Ostriker}, J.~P., {et~al.} 2001, ApJ, 552, 473

\bibitem[{{David} {et~al.}(2001){David}, {Nulsen}, {McNamara}, {Forman},
  {Jones}, {Ponman}, {Robertson}, \& {Wise}}]{david01}
{David}, L.~P., {Nulsen}, P.~E.~J., {McNamara}, B.~R., {et~al.} 2001, ApJ, 557,
  546

\bibitem[{{David} {et~al.}(2011){David}, {O'Sullivan}, {Jones}, {Giacintucci},
  {Vrtilek}, {Raychaudhury}, {Nulsen}, {Forman}, {Sun}, \& {Donahue}}]{david11}
{David}, L.~P., {O'Sullivan}, E., {Jones}, C., {et~al.} 2011, ApJ, 728, 162

\bibitem[{{Fabbiano}(2006)}]{fabbiano06}
{Fabbiano}, G. 2006, ARA\&A, 44, 323

\bibitem[{{Fabian}(1994)}]{fabian94}
{Fabian}, A.~C. 1994, ARA\&A, 32, 277

\bibitem[{{Fabian}(2012)}]{fabian12}
---. 2012, ARA\&A, 50, 455

\bibitem[{{Fabian} \& {Nulsen}(1977)}]{fabian77}
{Fabian}, A.~C., \& {Nulsen}, P.~E.~J. 1977, MNRAS, 180, 479

\bibitem[{{Fields} {et~al.}(2014){Fields}, {Molaro}, \& {Sarkar}}]{fields14}
{Fields}, B.~D., {Molaro}, P., \& {Sarkar}, S. 2014, arXiv e-prints,
  arXiv:1412.1408

\bibitem[{{Forman} {et~al.}(1985){Forman}, {Jones}, \& {Tucker}}]{forman85}
{Forman}, W., {Jones}, C., \& {Tucker}, W. 1985, ApJ, 293, 102

\bibitem[{{Forman} {et~al.}(1979){Forman}, {Schwarz}, {Jones}, {Liller}, \&
  {Fabian}}]{forman79}
{Forman}, W., {Schwarz}, J., {Jones}, C., {Liller}, W., \& {Fabian}, A.~C.
  1979, ApJL, 234, L27

\bibitem[{{Fukazawa} {et~al.}(2006){Fukazawa}, {Botoya-Nonesa}, {Pu}, {Ohto},
  \& {Kawano}}]{fukazawa06}
{Fukazawa}, Y., {Botoya-Nonesa}, J.~G., {Pu}, J., {Ohto}, A., \& {Kawano}, N.
  2006, ApJ, 636, 698

\bibitem[{{Fukugita} \& {Peebles}(2006)}]{fukugita06}
{Fukugita}, M., \& {Peebles}, P.~J.~E. 2006, ApJ, 639, 590

\bibitem[{{Gilfanov}(2004)}]{gilfanov04}
{Gilfanov}, M. 2004, MNRAS, 349, 146

\bibitem[{{Giodini} {et~al.}(2009){Giodini}, {Pierini}, {Finoguenov}, {Pratt},
  {Boehringer}, {Leauthaud}, {Guzzo}, {Aussel}, {Bolzonella}, {Capak}, {Elvis},
  {Hasinger}, {Ilbert}, {Kartaltepe}, {Koekemoer}, {Lilly}, {Massey},
  {McCracken}, {Rhodes}, {Salvato}, {Sanders}, {Scoville}, {Sasaki}, {Smolcic},
  {Taniguchi}, {Thompson}, \& {COSMOS Collaboration}}]{giodini09}
{Giodini}, S., {Pierini}, D., {Finoguenov}, A., {et~al.} 2009, ApJ, 703, 982

\bibitem[{{Goulding} {et~al.}(2016){Goulding}, {Greene}, {Ma}, {Veale},
  {Bogdan}, {Nyland}, {Blakeslee}, {McConnell}, \& {Thomas}}]{goulding16}
{Goulding}, A.~D., {Greene}, J.~E., {Ma}, C.-P., {et~al.} 2016, ApJ, 826, 167

\bibitem[{{Greco} {et~al.}(2015){Greco}, {Hill}, {Spergel}, \&
  {Battaglia}}]{greco15}
{Greco}, J.~P., {Hill}, J.~C., {Spergel}, D.~N., \& {Battaglia}, N. 2015, ApJ,
  808, 151

\bibitem[{{Grevesse} \& {Sauval}(1998)}]{grevesse98}
{Grevesse}, N., \& {Sauval}, A.~J. 1998, SSR, 85, 161

\bibitem[{{Gupta} {et~al.}(2017){Gupta}, {Mathur}, \& {Krongold}}]{gupta17}
{Gupta}, A., {Mathur}, S., \& {Krongold}, Y. 2017, ApJ, 836, 243

\bibitem[{{Gupta} {et~al.}(2012){Gupta}, {Mathur}, {Krongold}, {Nicastro}, \&
  {Galeazzi}}]{gupta12}
{Gupta}, A., {Mathur}, S., {Krongold}, Y., {Nicastro}, F., \& {Galeazzi}, M.
  2012, ApJL, 756, L8

\bibitem[{{Humphrey} {et~al.}(2006){Humphrey}, {Buote}, {Gastaldello},
  {Zappacosta}, {Bullock}, {Brighenti}, \& {Mathews}}]{humphrey06}
{Humphrey}, P.~J., {Buote}, D.~A., {Gastaldello}, F., {et~al.} 2006, ApJ, 646,
  899

\bibitem[{{Jones} {et~al.}(2002){Jones}, {Forman}, {Vikhlinin}, {Markevitch},
  {David}, {Warmflash}, {Murray}, \& {Nulsen}}]{jones02}
{Jones}, C., {Forman}, W., {Vikhlinin}, A., {et~al.} 2002, ApJL, 567, L115

\bibitem[{{Kelly} {et~al.}(2021){Kelly}, {Jenkins}, \& {Frenk}}]{kelly21}
{Kelly}, A.~J., {Jenkins}, A., \& {Frenk}, C.~S. 2021, MNRAS, 502, 2934

\bibitem[{{Kim} \& {Fabbiano}(2013)}]{kim13}
{Kim}, D.-W., \& {Fabbiano}, G. 2013, ApJ, 776, 116

\bibitem[{{Kov{\'a}cs} {et~al.}(2019){Kov{\'a}cs}, {Bogd{\'a}n}, {Smith},
  {Kraft}, \& {Forman}}]{kovacs19}
{Kov{\'a}cs}, O.~E., {Bogd{\'a}n}, {\'A}., {Smith}, R.~K., {Kraft}, R.~P., \&
  {Forman}, W.~R. 2019, ApJ, 872, 83

\bibitem[{{Li} {et~al.}(2018){Li}, {Bregman}, {Wang}, {Crain}, \&
  {Anderson}}]{li18}
{Li}, J.-T., {Bregman}, J.~N., {Wang}, Q.~D., {Crain}, R.~A., \& {Anderson},
  M.~E. 2018, ApJL, 855, L24

\bibitem[{{Li} {et~al.}(2017){Li}, {Bregman}, {Wang}, {Crain}, {Anderson}, \&
  {Zhang}}]{li17}
{Li}, J.-T., {Bregman}, J.~N., {Wang}, Q.~D., {et~al.} 2017, ApJS, 233, 20

\bibitem[{{Li} {et~al.}(2014){Li}, {Crain}, \& {Wang}}]{li14}
{Li}, J.-T., {Crain}, R.~A., \& {Wang}, Q.~D. 2014, MNRAS, 440, 859

\bibitem[{{Li} \& {Wang}(2013)}]{li13}
{Li}, J.-T., \& {Wang}, Q.~D. 2013, MNRAS, 435, 3071

\bibitem[{{Lim} {et~al.}(2021){Lim}, {Barnes}, {Vogelsberger}, {Mo}, {Nelson},
  {Pillepich}, {Dolag}, \& {Marinacci}}]{lim21}
{Lim}, S.~H., {Barnes}, D., {Vogelsberger}, M., {et~al.} 2021, MNRAS, 504, 5131

\bibitem[{{Mannucci} {et~al.}(2005){Mannucci}, {Della Valle}, {Panagia},
  {Cappellaro}, {Cresci}, {Maiolino}, {Petrosian}, \& {Turatto}}]{mannucci05}
{Mannucci}, F., {Della Valle}, M., {Panagia}, N., {et~al.} 2005, A\&A, 433, 807

\bibitem[{{Mathews}(1978)}]{mathews78}
{Mathews}, W.~G. 1978, ApJ, 219, 413

\bibitem[{{Mathews}(1990)}]{mathews90}
---. 1990, ApJ, 354, 468

\bibitem[{{Mathews} \& {Brighenti}(2003)}]{mathews03}
{Mathews}, W.~G., \& {Brighenti}, F. 2003, ARA\&A, 41, 191

\bibitem[{{McGaugh} {et~al.}(2000){McGaugh}, {Schombert}, {Bothun}, \& {de
  Blok}}]{mcgaugh00}
{McGaugh}, S.~S., {Schombert}, J.~M., {Bothun}, G.~D., \& {de Blok}, W.~J.~G.
  2000, ApJL, 533, L99

\bibitem[{{McNamara} \& {Nulsen}(2007)}]{mcnamara07}
{McNamara}, B.~R., \& {Nulsen}, P.~E.~J. 2007, ARA\&A, 45, 117

\bibitem[{{McNamara} \& {Nulsen}(2012)}]{mcnamara12}
---. 2012, NJPh, 14, 055023

\bibitem[{{Mernier} {et~al.}(2020){Mernier}, {Werner}, {Lakhchaura}, {de Plaa},
  {Gu}, {Kaastra}, {Mao}, {Simionescu}, \& {Urdampilleta}}]{mernier20}
{Mernier}, F., {Werner}, N., {Lakhchaura}, K., {et~al.} 2020, Astronomische
  Nachrichten, 341, 203

\bibitem[{{Nagai} {et~al.}(2007){Nagai}, {Vikhlinin}, \& {Kravtsov}}]{nagai07}
{Nagai}, D., {Vikhlinin}, A., \& {Kravtsov}, A.~V. 2007, ApJ, 655, 98

\bibitem[{{Navarro} {et~al.}(1997){Navarro}, {Frenk}, \& {White}}]{navarro97}
{Navarro}, J.~F., {Frenk}, C.~S., \& {White}, S. D.~M. 1997, ApJ, 490, 493

\bibitem[{{Oppenheimer} \& {Dav{\'e}}(2006)}]{oppenheimer2006}
{Oppenheimer}, B.~D., \& {Dav{\'e}}, R. 2006, MNRAS, 373, 1265

\bibitem[{{Oppenheimer} {et~al.}(2020){Oppenheimer}, {Bogd{\'a}n}, {Crain},
  {ZuHone}, {Forman}, {Schaye}, {Wijers}, {Davies}, {Jones}, {Kraft}, \&
  {Ghirardini}}]{oppenheimer20}
{Oppenheimer}, B.~D., {Bogd{\'a}n}, {\'A}., {Crain}, R.~A., {et~al.} 2020,
  ApJL, 893, L24

\bibitem[{{O'Sullivan} {et~al.}(2001){O'Sullivan}, {Forbes}, \&
  {Ponman}}]{osullivan01}
{O'Sullivan}, E., {Forbes}, D.~A., \& {Ponman}, T.~J. 2001, MNRAS, 328, 461

\bibitem[{{O'Sullivan} {et~al.}(2003){O'Sullivan}, {Ponman}, \&
  {Collins}}]{osullivan03}
{O'Sullivan}, E., {Ponman}, T.~J., \& {Collins}, R.~S. 2003, MNRAS, 340, 1375

\bibitem[{{Peterson} \& {Fabian}(2006)}]{peterson06}
{Peterson}, J.~R., \& {Fabian}, A.~C. 2006, PhR, 427, 1

\bibitem[{{Peterson} {et~al.}(2003){Peterson}, {Kahn}, {Paerels}, {Kaastra},
  {Tamura}, {Bleeker}, {Ferrigno}, \& {Jernigan}}]{peterson03}
{Peterson}, J.~R., {Kahn}, S.~M., {Paerels}, F.~B.~S., {et~al.} 2003, ApJ, 590,
  207

\bibitem[{{Planck Collaboration} {et~al.}(2013){Planck Collaboration}, {Ade},
  {Aghanim}, {Arnaud}, {Ashdown}, {Atrio-Barandela}, {Aumont}, {Baccigalupi},
  {Balbi}, {Banday}, {Barreiro}, {Barrena}, {Bartlett}, {Battaner}, {Benabed},
  {Bernard}, {Bersanelli}, {Bikmaev}, {Bock}, {B{\"o}hringer}, {Bonaldi},
  {Bond}, {Borrill}, {Bouchet}, {Bourdin}, {Burenin}, {Burigana}, {Butler},
  {Cabella}, {Chamballu}, {Chary}, {Chiang}, {Chon}, {Christensen}, {Clements},
  {Colafrancesco}, {Colombi}, {Colombo}, {Comis}, {Coulais}, {Crill},
  {Cuttaia}, {Da Silva}, {Dahle}, {Davis}, {de Bernardis}, {de Gasperis}, {de
  Rosa}, {de Zotti}, {Delabrouille}, {D{\'e}mocl{\`e}s}, {Diego}, {Dole},
  {Donzelli}, {Dor{\'e}}, {Douspis}, {Dupac}, {Efstathiou}, {En{\ss}lin},
  {Finelli}, {Flores-Cacho}, {Forni}, {Frailis}, {Franceschi}, {Frommert},
  {Galeotta}, {Ganga}, {G{\'e}nova-Santos}, {Giard}, {Giraud-H{\'e}raud},
  {Gonz{\'a}lez-Nuevo}, {G{\'o}rski}, {Gregorio}, {Gruppuso}, {Hansen},
  {Harrison}, {Hern{\'a}ndez-Monteagudo}, {Herranz}, {Hildebrandt}, {Hivon},
  {Hobson}, {Holmes}, {Hornstrup}, {Hovest}, {Huffenberger}, {Hurier}, {Jaffe},
  {Jaffe}, {Jones}, {Juvela}, {Keih{\"a}nen}, {Keskitalo}, {Khamitov},
  {Kisner}, {Kneissl}, {Knoche}, {Kunz}, {Kurki-Suonio}, {L{\"a}hteenm{\"a}ki},
  {Lamarre}, {Lasenby}, {Lawrence}, {Le Jeune}, {Leonardi}, {Lilje},
  {Linden-V{\o}rnle}, {L{\'o}pez-Caniego}, {Lubin}, {Luzzi},
  {Mac{\'\i}as-P{\'e}rez}, {MacTavish}, {Maffei}, {Maino}, {Mandolesi},
  {Maris}, {Marleau}, {Marshall}, {Mart{\'\i}nez-Gonz{\'a}lez}, {Masi},
  {Massardi}, {Matarrese}, {Mazzotta}, {Mei}, {Melchiorri}, {Melin}, {Mendes},
  {Mennella}, {Mitra}, {Miville-Desch{\^e}nes}, {Moneti}, {Montier},
  {Morgante}, {Mortlock}, {Munshi}, {Murphy}, {Naselsky}, {Nati}, {Natoli},
  {N{\o}rgaard-Nielsen}, {Noviello}, {Novikov}, {Novikov}, {Osborne},
  {Oxborrow}, {Pajot}, {Paoletti}, {Perotto}, {Perrotta}, {Piacentini}, {Piat},
  {Pierpaoli}, {Piffaretti}, {Plaszczynski}, {Pointecouteau}, {Polenta},
  {Popa}, {Poutanen}, {Pratt}, {Prunet}, {Puget}, {Rachen}, {Rebolo},
  {Reinecke}, {Remazeilles}, {Renault}, {Ricciardi}, {Ristorcelli}, {Rocha},
  {Roman}, {Rosset}, {Rossetti}, {Rubi{\~n}o-Mart{\'\i}n}, {Rusholme},
  {Sandri}, {Savini}, {Scott}, {Spencer}, {Starck}, {Stolyarov}, {Sudiwala},
  {Sunyaev}, {Sutton}, {Suur-Uski}, {Sygnet}, {Tauber}, {Terenzi},
  {Toffolatti}, {Tomasi}, {Tristram}, {Valenziano}, {Van Tent}, {Vielva},
  {Villa}, {Vittorio}, {Wade}, {Wandelt}, {Wang}, {Welikala}, {Weller},
  {White}, {White}, {Yvon}, {Zacchei}, \& {Zonca}}]{planck13}
{Planck Collaboration}, {Ade}, P.~A.~R., {Aghanim}, N., {et~al.} 2013, A\&A,
  557, A52

\bibitem[{{Planck Collaboration} {et~al.}(2016){Planck Collaboration}, {Ade},
  {Aghanim}, {Arnaud}, {Ashdown}, {Aumont}, {Baccigalupi}, {Banday},
  {Barreiro}, {Bartlett}, {Bartolo}, {Battaner}, {Battye}, {Benabed},
  {Beno{\^\i}t}, {Benoit-L{\'e}vy}, {Bernard}, {Bersanelli}, {Bielewicz},
  {Bock}, {Bonaldi}, {Bonavera}, {Bond}, {Borrill}, {Bouchet}, {Boulanger},
  {Bucher}, {Burigana}, {Butler}, {Calabrese}, {Cardoso}, {Catalano},
  {Challinor}, {Chamballu}, {Chary}, {Chiang}, {Chluba}, {Christensen},
  {Church}, {Clements}, {Colombi}, {Colombo}, {Combet}, {Coulais}, {Crill},
  {Curto}, {Cuttaia}, {Danese}, {Davies}, {Davis}, {de Bernardis}, {de Rosa},
  {de Zotti}, {Delabrouille}, {D{\'e}sert}, {Di Valentino}, {Dickinson},
  {Diego}, {Dolag}, {Dole}, {Donzelli}, {Dor{\'e}}, {Douspis}, {Ducout},
  {Dunkley}, {Dupac}, {Efstathiou}, {Elsner}, {En{\ss}lin}, {Eriksen},
  {Farhang}, {Fergusson}, {Finelli}, {Forni}, {Frailis}, {Fraisse},
  {Franceschi}, {Frejsel}, {Galeotta}, {Galli}, {Ganga}, {Gauthier}, {Gerbino},
  {Ghosh}, {Giard}, {Giraud-H{\'e}raud}, {Giusarma}, {Gjerl{\o}w},
  {Gonz{\'a}lez-Nuevo}, {G{\'o}rski}, {Gratton}, {Gregorio}, {Gruppuso},
  {Gudmundsson}, {Hamann}, {Hansen}, {Hanson}, {Harrison}, {Helou},
  {Henrot-Versill{\'e}}, {Hern{\'a}ndez-Monteagudo}, {Herranz}, {Hildebrandt},
  {Hivon}, {Hobson}, {Holmes}, {Hornstrup}, {Hovest}, {Huang}, {Huffenberger},
  {Hurier}, {Jaffe}, {Jaffe}, {Jones}, {Juvela}, {Keih{\"a}nen}, {Keskitalo},
  {Kisner}, {Kneissl}, {Knoche}, {Knox}, {Kunz}, {Kurki-Suonio}, {Lagache},
  {L{\"a}hteenm{\"a}ki}, {Lamarre}, {Lasenby}, {Lattanzi}, {Lawrence}, {Leahy},
  {Leonardi}, {Lesgourgues}, {Levrier}, {Lewis}, {Liguori}, {Lilje},
  {Linden-V{\o}rnle}, {L{\'o}pez-Caniego}, {Lubin}, {Mac{\'\i}as-P{\'e}rez},
  {Maggio}, {Maino}, {Mandolesi}, {Mangilli}, {Marchini}, {Maris}, {Martin},
  {Martinelli}, {Mart{\'\i}nez-Gonz{\'a}lez}, {Masi}, {Matarrese}, {McGehee},
  {Meinhold}, {Melchiorri}, {Melin}, {Mendes}, {Mennella}, {Migliaccio},
  {Millea}, {Mitra}, {Miville-Desch{\^e}nes}, {Moneti}, {Montier}, {Morgante},
  {Mortlock}, {Moss}, {Munshi}, {Murphy}, {Naselsky}, {Nati}, {Natoli},
  {Netterfield}, {N{\o}rgaard-Nielsen}, {Noviello}, {Novikov}, {Novikov},
  {Oxborrow}, {Paci}, {Pagano}, {Pajot}, {Paladini}, {Paoletti}, {Partridge},
  {Pasian}, {Patanchon}, {Pearson}, {Perdereau}, {Perotto}, {Perrotta},
  {Pettorino}, {Piacentini}, {Piat}, {Pierpaoli}, {Pietrobon}, {Plaszczynski},
  {Pointecouteau}, {Polenta}, {Popa}, {Pratt}, {Pr{\'e}zeau}, {Prunet},
  {Puget}, {Rachen}, {Reach}, {Rebolo}, {Reinecke}, {Remazeilles}, {Renault},
  {Renzi}, {Ristorcelli}, {Rocha}, {Rosset}, {Rossetti}, {Roudier},
  {Rouill{\'e} d'Orfeuil}, {Rowan-Robinson}, {Rubi{\~n}o-Mart{\'\i}n},
  {Rusholme}, {Said}, {Salvatelli}, {Salvati}, {Sandri}, {Santos},
  {Savelainen}, {Savini}, {Scott}, {Seiffert}, {Serra}, {Shellard}, {Spencer},
  {Spinelli}, {Stolyarov}, {Stompor}, {Sudiwala}, {Sunyaev}, {Sutton},
  {Suur-Uski}, {Sygnet}, {Tauber}, {Terenzi}, {Toffolatti}, {Tomasi},
  {Tristram}, {Trombetti}, {Tucci}, {Tuovinen}, {T{\"u}rler}, {Umana},
  {Valenziano}, {Valiviita}, {Van Tent}, {Vielva}, {Villa}, {Wade}, {Wandelt},
  {Wehus}, {White}, {White}, {Wilkinson}, {Yvon}, {Zacchei}, \&
  {Zonca}}]{planck16}
---. 2016, A\&A, 594, A13

\bibitem[{{Randall} {et~al.}(2011){Randall}, {Forman}, {Giacintucci}, {Nulsen},
  {Sun}, {Jones}, {Churazov}, {David}, {Kraft}, {Donahue}, {Blanton},
  {Simionescu}, \& {Werner}}]{randall11}
{Randall}, S.~W., {Forman}, W.~R., {Giacintucci}, S., {et~al.} 2011, ApJ, 726,
  86

\bibitem[{{Rasmussen} {et~al.}(2009){Rasmussen}, {Sommer-Larsen}, {Pedersen},
  {Toft}, {Benson}, {Bower}, \& {Grove}}]{rasmussen09}
{Rasmussen}, J., {Sommer-Larsen}, J., {Pedersen}, K., {et~al.} 2009, ApJ, 697,
  79

\bibitem[{{Revnivtsev} {et~al.}(2008){Revnivtsev}, {Churazov}, {Sazonov},
  {Forman}, \& {Jones}}]{revnivtsev08}
{Revnivtsev}, M., {Churazov}, E., {Sazonov}, S., {Forman}, W., \& {Jones}, C.
  2008, A\&A, 490, 37

\bibitem[{{Revnivtsev} {et~al.}(2006){Revnivtsev}, {Sazonov}, {Gilfanov},
  {Churazov}, \& {Sunyaev}}]{revnivtsev06}
{Revnivtsev}, M., {Sazonov}, S., {Gilfanov}, M., {Churazov}, E., \& {Sunyaev},
  R. 2006, A\&A, 452, 169

\bibitem[{{Sazonov} {et~al.}(2006){Sazonov}, {Revnivtsev}, {Gilfanov},
  {Churazov}, \& {Sunyaev}}]{sazonov06}
{Sazonov}, S., {Revnivtsev}, M., {Gilfanov}, M., {Churazov}, E., \& {Sunyaev},
  R. 2006, A\&A, 450, 117

\bibitem[{{Shull} {et~al.}(2012){Shull}, {Smith}, \& {Danforth}}]{shull2012}
{Shull}, J.~M., {Smith}, B.~D., \& {Danforth}, C.~W. 2012, ApJ, 759, 23

\bibitem[{{Spitzer}(1956)}]{spitzer56}
{Spitzer}, Lyman, J. 1956, ApJ, 124, 20

\bibitem[{{Strickland} {et~al.}(2004){Strickland}, {Heckman}, {Colbert},
  {Hoopes}, \& {Weaver}}]{strickland04}
{Strickland}, D.~K., {Heckman}, T.~M., {Colbert}, E. J.~M., {Hoopes}, C.~G., \&
  {Weaver}, K.~A. 2004, ApJS, 151, 193

\bibitem[{{Suresh} {et~al.}(2017){Suresh}, {Rubin}, {Kannan}, {Werk},
  {Hernquist}, \& {Vogelsberger}}]{suresh2017}
{Suresh}, J., {Rubin}, K. H.~R., {Kannan}, R., {et~al.} 2017, MNRAS, 465, 2966

\bibitem[{{Theuns} {et~al.}(2002){Theuns}, {Viel}, {Kay}, {Schaye}, {Carswell},
  \& {Tzanavaris}}]{theuns2002}
{Theuns}, T., {Viel}, M., {Kay}, S., {et~al.} 2002, ApJL, 578, L5

\bibitem[{{Toft} {et~al.}(2002){Toft}, {Rasmussen}, {Sommer-Larsen}, \&
  {Pedersen}}]{toft02}
{Toft}, S., {Rasmussen}, J., {Sommer-Larsen}, J., \& {Pedersen}, K. 2002,
  MNRAS, 335, 799

\bibitem[{{Trinchieri} \& {Fabbiano}(1985)}]{trinchieri85}
{Trinchieri}, G., \& {Fabbiano}, G. 1985, ApJ, 296, 447

\bibitem[{{T{\"u}llmann} {et~al.}(2006){T{\"u}llmann}, {Pietsch}, {Rossa},
  {Breitschwerdt}, \& {Dettmar}}]{tullmann06}
{T{\"u}llmann}, R., {Pietsch}, W., {Rossa}, J., {Breitschwerdt}, D., \&
  {Dettmar}, R.~J. 2006, A\&A, 448, 43

\bibitem[{{van de Voort} {et~al.}(2019){van de Voort}, {Springel}, {Mandelker},
  {van den Bosch}, \& {Pakmor}}]{vandevoort2019}
{van de Voort}, F., {Springel}, V., {Mandelker}, N., {van den Bosch}, F.~C., \&
  {Pakmor}, R. 2019, MNRAS, 482, L85

\bibitem[{{White} \& {Frenk}(1991)}]{white91}
{White}, S. D.~M., \& {Frenk}, C.~S. 1991, ApJ, 379, 52

\bibitem[{{White} \& {Rees}(1978)}]{white78}
{White}, S.~D.~M., \& {Rees}, M.~J. 1978, MNRAS, 183, 341

\bibitem[{{Wijers} {et~al.}(2020){Wijers}, {Schaye}, \&
  {Oppenheimer}}]{wijers2020}
{Wijers}, N.~A., {Schaye}, J., \& {Oppenheimer}, B.~D. 2020, MNRAS, 498, 574

\bibitem[{{Yao} {et~al.}(2010){Yao}, {Wang}, {Penton}, {Tripp}, {Shull}, \&
  {Stocke}}]{yao10}
{Yao}, Y., {Wang}, Q.~D., {Penton}, S.~V., {et~al.} 2010, ApJ, 716, 1514

\end{thebibliography}

\end{document}